%% file: spi.tex
\documentclass[usenatbib,useAMS]{mn2e} %

\usepackage{times}
\usepackage{graphicx} %
\usepackage{color,array,colortbl,longtable}%
\usepackage{amsmath,amssymb,amscd} %
\newcommand{\bc}{\begin{center}}%
\newcommand{\ec}{\end{center}}%
\newcommand{\be}{\begin{eqnarray}} %
\newcommand{\ee}{\end{eqnarray}} %
\newcommand{\intgr}{{\it INTEGRAL}} %
\newcommand{\gr}{$\gamma$-ray} %

\IfFileExists{srcltx.sty}{\usepackage[active]{srcltx}}

\usepackage{bm,paralist,xspace,url} %
\newcommand{\eprint}[1]{\href{http://arxiv.org/abs/#1}{#1}}

\newcommand{\adsurl}[1]{\href{#1}{ADS}}

\usepackage[colorlinks=false,bookmarks=false]{hyperref}

\IfFileExists{hypernat.sty}{\usepackage{hypernat}}

\newcommand{\fov}{{\mathrm{FoV}}} 
\newcommand{\pcfov}{{\textsc{pcfov}}} 
\newcommand{\fcfov}{{\textsc{fcfov}}} 
\newcommand{\eff}{{\mathrm{eff}}} 
\renewcommand{\min}{{\text{min}}} 
\renewcommand{\max}{{\text{max}}} 
\newcommand{\vir}{{\mathrm{vir}}} 
\newcommand{\nfw}{{\textsc{nfw}}} 
\newcommand{\iso}{{\mathrm{iso}}} 
\newcommand{\cm}{\:\mathrm{cm}} 
\newcommand{\dm}{{\textsc{dm}}} 
\newcommand{\mdm}{\ensuremath{M_\textsc{dm}}\xspace} 
\newcommand{\numsm}{$\nu$MSM\xspace} 
\newcommand{\ev}{\:\mathrm{eV}} 
\newcommand{\kev}{\:\mathrm{keV}} 
\newcommand{\keV}{\:\mathrm{keV}} 
\newcommand{\mev}{\:\mathrm{MeV}} 
\newcommand{\gev}{\:\mathrm{GeV}} 
\newcommand{\kpc}{\:\mathrm{kpc}} 
\newcommand{\mpc}{\:\mathrm{Mpc}} 
\newcommand{\CS}{\mathcal{S}} 

\title[Constraining DM properties with SPI]{\hfill \hbox{\small CERN-PH-TH/2007-202}\\[\baselineskip]
Constraining DM properties with SPI} %

\author[A.~Boyarsky et al.]{Alexey~Boyarsky$^{1}$\thanks{On leave of absence
    from Bogolyubov Institute for Theoretical Physics, Kiev, Ukraine},
  Denys~Malyshev$^{2,3}$, Andrey~Neronov$^4$\footnotemark[1],
  Oleg~Ruchayskiy$^5$\\ %
  \small
  $^1$ CERN, PH/TH, CH-1211 Geneve 23,  Switzerland \\
  $^2$ Bogolyubov Institute for Theoretical Physics, Kiev,
  03780, Ukraine\\
  $^3$ Dublin Institute for Advanced Studies, 31 Fitzwilliam Place, Dublin 2, Ireland\\
  $^4$ INTEGRAL Science Data Center, Chemin d'\'Ecogia 16, 1290
  Versoix, Switzerland \\
  {}Geneva Observatory, 51 ch. des Maillettes, CH-1290
  Sauverny, Switzerland\\
  $^5$ \'Ecole Polytechnique F\'ed\'erale de Lausanne, Institute of
  Theoretical Physics,\\ FSB/ITP/LPPC, BSP 720, CH-1015, Lausanne,
  Switzerland}
\date{}%

\begin{document}
\label{firstpage}

\maketitle

\begin{abstract}
Using the high-resolution spectrometer SPI on board the \emph{International 
Gamma-Ray Astrophysics Laboratory (INTEGRAL)}, we search for a spectral line
 produced by a dark matter (DM) particle with a mass in the range $40 keV < M_{DM} < 14 MeV$,
 decaying in the DM halo of the Milky Way. To distinguish the DM decay line from numerous 
instrumental lines found in the SPI background spectrum, we study the dependence of the 
intensity of the line signal on the offset of the SPI pointing from the direction toward 
the Galactic Centre. After a critical analysis of the uncertainties of the DM density profile
 in the inner Galaxy, we find that the intensity of the DM decay line should decrease by at
 least a factor of 3 when the offset from the Galactic Centre increases from $0^\circ$ to
 $180^\circ$. We find that such a pronounced variation of the line flux across the sky is not
 observed for any line, detected with a significance higher than $3\sigma$ in the SPI
 background spectrum. Possible DM decay origin is not ruled out only for the unidentified
 spectral lines, having low ($\sim 3\sigma$) significance or coinciding in position with the
instrumental ones. In the energy interval from 20 keV to 7 MeV, we derive restrictions on
the DM decay line flux, implied by the (non-)detection of the DM decay line. For a particular
DM candidate, the sterile neutrino of mass $M_{DM}$, we derive a bound on the mixing angle.

\end{abstract}
\pagerange{\pageref{firstpage}--\pageref{lastpage}} \pubyear{2007}
\section{Introduction}

\subsubsection*{Dark matter in the Universe}

There is a vast body of evidence, suggesting that the large fraction of
matter in the Universe exists in the form of the \emph{Dark matter}
\emph{(DM)}. However, while the total density of the DM is measured with a
very high precision ($\Omega_{\rm DM}h^2 =
0.105^{+0.007}_{-0.009}$,~\citealt{WMAP3}), little is known about its
properties apart from this.
The possibility that the DM is composed of the Standard Model (SM) particles
has been ruled out for a long time already.  Indeed, the DM cannot be made out
of baryons, as producing such an amount of baryonic matter would require
drastic modifications of the scenario of the Big Bang nucleosynthesis (BBN),
which otherwise successfully describes the abundance of light elements~(see
for example ~\citealt{Dar:95}). Recent microlensing experiments rule out the
possibility that another type of baryonic DM -- massive compact halo objects (MACHOs) -- constitute dominant fraction of mass in the
halo~\citep{MACHO:00,EROS:00,OGLE:98}.
The only non-baryonic DM candidate in the SM candidates -- (left-handed)
neutrino -- is ruled out from the large scale structure (LSS)
considerations~\citep[see e.g.][]{Bond:80,Hannestad:03,Crotty:04}.

What are the properties of a successful DM candidate?  First of all, this
particle should be massive.  Many extensions of the SM present the DM
candidates with the masses ranging from $\sim 10^{-10}\ev$~(massive gravitons,
\citealt{Dubovsky:04}) and $\sim 10^{-6}$~eV (axions) to hundreds of GeV
(WIMPs) and even to $10^{13}$~GeV
\citep[WIMPZILLA,][]{Kuzmin:98,Kuzmin:99,Chung:98}. For a review of particle
physics DM candidates see e.g.~\cite{Bergstrom:00,Bertone:04,Carr:06}.

Secondly, there should exist mechanisms of DM production with the correct
abundances. The production mechanism in particular determines the velocity
distribution of particles in the early Universe. This velocity distribution
can, in principle, be probed experimentally.  Namely, if during the structure
formation epoch the DM particles have velocities, comparable to the speed of
sound in the baryon-photon plasma, they ``erase'' density fluctuations at
scales, smaller than the distance, they have traveled (called the \emph{free-streaming
  length}).  To differentiate various models in accordance with this property,
the DM candidates with the negligible velocity dispersion (and,
correspondingly, free-streaming) are called \emph{cold} DM (CDM), while those
with the free-streaming of the order of $\sim 1\mpc$ are considered to be
\emph{warm} (WDM).\footnote{The left-handed neutrino would represent
  \emph{hot} DM in this terminology, i.e.  the DM with the free-streaming
  length $\gg 1$~Mpc.}
It is possible to constrain the free-streaming length of a particular DM
candidate by probing the structure of the Universe at galaxy-size scales.
This can be done through the analysis of the Lyman-$\alpha$ forest
data~\citep{Hui:97}.  Lyman-$\alpha$ analysis puts an upper bound on the
free-streaming of the DM
particles~\citep{Hansen:01,Viel:05,Seljak:06,Viel:06,Viel:07}. It should be
noted however that currently existing interpretation of the Lyman-$\alpha$
data is model-dependent, as, apart from a number of astrophysical assumptions
(see~\citealt{Hui:97}) and complicated hydrodynamic simulations, it relies on
{\it a priori} assumptions about the velocity distribution of the DM
particles.

A way to differentiate between CDM and WDM models would be to compare the
numerical simulations of the DM distribution in the Milky Way-type galaxies
with the actual observations.  However, the resolution of the N-body
simulations is not yet sufficient to answer the questions about e.g. the DM
density profiles in dwarf satellite galaxies. Moreover, most of the
simulations include only collisionless DM particles, and do not model the
baryons and their feedback on the galaxy structure formation.  These problems
are not solved even for the CDM simulations, and WDM simulations have
additional serious difficulties.  From an observational point of view, it has been
argued for some time already that there is a discrepancy between CDM
simulations and observations (see
e.g.~\citealt{Moore:94,Moore:99,Klypin:99,Bode:00,Avila-Reese:01,Goerdt:06})
It has been claimed recently that a number of recent observations of dwarf
satellite galaxies of the Milky way and Andromeda galaxy seem to indicate the
existence of the smallest scale at which the DM
exists~\citep{Gilmore:06,Gilmore:07a,Gilmore:07b,Koposov:07}. However, this
statement and the interpretation of the observations are still subject to
debate~\citep{Klimentowski:06,Penarrubia:07,Strigari:07,Simon:07}.  Therefore
it is too early to say what kind of DM models is favoured by comparing
simulations and observations.


Usually it is also necessary for the DM candidate to be stable. For the most
popular DM candidate -- weakly interacting massive particles (WIMPs), this is
related to the fact that the particles of $\sim$ electroweak mass, having weak
strength interaction with SM matter (required to produce the correct amount of
DM), would decay too fast and would not be ``dark''.  If, however, the DM
particle interacts with the SM more weakly than WIMPs, it could well have a
finite (although cosmologically long) life time.

There exist several unstable (decaying) DM candidates e.g.
gravitino~\citep{Borgani:96,Baltz:01,Roszkowski:04,Cerdeno:05,Cembranos:06,Lola:07}.
In this paper we will concentrate mainly on one candidate, the sterile neutrino
(although our results will be applicable for any type of decaying DM).
Constraints on the decaying DM were analyzed
in~\cite{DeRujula:80,Berezhiani:87,Doroshkevich:89,Berezhiani:90a,Berezhiani:90b,Bertone:07,Zhang:07}
(see also the book by~\citealt{Khlopov:97}).

\subsubsection*{Sterile neutrino DM}
\label{sec:sterile-neutrino-dm}

It was noticed long ago that the right-handed (or as it is often called
\emph{sterile}) neutrino with the mass in the keV range would represent a
viable DM candidate~\citep{Dodelson:93}. Such a neutrino would interact with
the rest of the matter only via the quadratic mixing with left-handed
(\emph{active}) neutrinos and therefore (although not stable) could have
cosmologically long life-time. At the same time, it could be produced in the
early Universe with the correct
abundances~\citep{Dodelson:93,Shi:98,Shaposhnikov:06}.  One of the decay
channels of the unstable sterile neutrinos includes emission of photons of the
energy equal to half of the sterile neutrino rest energy.  This potentially
provides a possibility to observe the decays of DM sterile neutrinos via
detection of a characteristic spectral line in the 
spectra of astrophysical objects with large DM concentration.

Recently this DM candidate has attracted much attention (see
e.g.~\citet{Shaposhnikov:07a} and references therein).  It was found that a
very modest and natural extension of the SM by 3 right-handed neutrinos
(making the SM more symmetric as all SM fermions, including neutrino, would
have now their left and right handed counterparts) provided a viable extension
of the theory, capable of solving several ``beyond the SM'' problems. First of
all, such an extension makes neutrinos massive and thus perhaps provides the
simplest and the most natural explanation of the phenomenon of ``neutrino
oscillations''~(see e.g.~\cite{Fogli:05,Strumia:06,Giunti:06} for reviews).
The smallness of neutrino masses in this model (called \numsm
in~\citealt{Asaka:05b}) is achieved by the usual see-saw mechanism with
Majorana masses of right-handed neutrinos being below electroweak
scale.\footnote{The fact that the \numsm does not introduce any new scale
  above the electroweak one, makes this theory especially appealing from the
  point of view of its experimental verification/falsification.}

Secondly, if two heavier sterile neutrinos ($N_2$ and $N_3$) are almost
degenerate in mass and have their masses between $\mathcal{O}(100)\mev$ and
$\mathcal{O}(20)\gev$, the \numsm provides the mechanism of generating the
baryon asymmetry of the Universe. Thirdly, the lightest sterile neutrino $N_1$
can have arbitrary mass and arbitrarily weak coupling with the (active)
neutrino sector. At the same time, it can be produced in the early Universe in
the correct amounts.  It represents therefore the DM particle in the \numsm.
Thus, altogether the \numsm represents (arguably) the simplest extension of
the SM, capable of explaining three important questions: origin and smallness
of neutrino masses, baryon asymmetry in the Universe and the existence of the
DM.

\subsubsection*{Existing restrictions on sterile neutrino DM parameters.}

What are the current restrictions on parameters (mass and \emph{mixing}) of
sterile neutrino DM? First of all sterile neutrino mass should satisfy the
universal Tremaine-Gunn lower bound:\footnote{In its simplest form the
  Tremaine-Gunn bound comes from the fact that for the fermions there is a
  maximal density in the phase space~\citep{Tremaine:79,Dalcanton:00} and
  therefore the observed phase-space density in various DM dominated systems
  should be less that this (mass dependent) bound.}  $\mdm\gtrsim
300-500\ev$.\footnote{A stronger lower bound from
  Ly-$\alpha$~\citep{Seljak:06,Viel:06,Viel:07} can be obtained in the case of
  the particular production mechanisms -- the Dodelson-Widrow
  scenario~\citep{Dodelson:93}. For other possible production
  mechanisms~\citep[e.g.][]{Shi:98,Shaposhnikov:06} the Ly-$\alpha$
  constraints should be reanalyzed.}

Next, as the sterile neutrino possesses the (two-body) radiative decay
channel: $N_1 \to \nu + \gamma$, the emitted photon would carry the energy
$E_\gamma = \mdm/2$. A large flux of such photons is expected from the large
concentrations of the DM sterile neutrinos, like galaxies or galaxy clusters.

Recently an extensive search of the DM decay line in the region of masses
$M_\dm \lesssim 20\kev$
was conducted, using the data of
\emph{Chandra}~\citep{Riemer:06,Boyarsky:06e,Abazajian:06b} and
\emph{XMM-Newton}~\citep{Boyarsky:05,Boyarsky:06b,Boyarsky:06c,Watson:06,Boyarsky:06d}.
The region of soft X-ray (down to energies $0.2$~keV) was explored
by~\citet{Boyarsky:06f} with the use of the wide field of view
spectrometer~\citep{McCammon:02}.  The non-observation of the DM decay line in
X-ray, combined with the first principles calculation of DM production in the
early Universe~\citep{Asaka:06c}, implies that the \citet{Dodelson:93} (DW)
scenario can work only if the sterile neutrino mass is below
4~keV~\citep{Boyarsky:07a}.  If one takes into account recent \emph{lower}
bound on the mass of sterile neutrino DM in the DW scenario $\mdm \ge
5.6\kev$~\citep{Viel:07}, it seems that the possibility that all the DM is
produced via DW scenario is ruled out~\citep{Boyarsky:07a}.  The possibility
that only fraction of the DM is produced via DW mechanism remains
open~\citep{Palazzo:07}.

There are other viable mechanisms of DM production, including e.g.  resonant
oscillation production in the presence of lepton asymmetries~\citep{Shi:98}.
Sterile neutrino DM can be produced by the decay of light inflaton
\citep{Shaposhnikov:06} or in a similar model with the different choice of
parameters~\citep{Kusenko:06a,Petraki:07}. These mechanisms are currently not
constrained and remain valid for DM particles with the masses in the keV range
and above.

The search for the DM decay line signal produced by sterile neutrinos with
masses above $\sim 20$~keV is complicated by the absence of the focusing
optics telescopes (similar to {\it Chandra} or {\it XMM-Newton}) in the hard
X-ray and $\gamma$-ray domain of the spectrum.  For example, the existing
restrictions in the $20-100$~keV mass range \citep{Boyarsky:05,Boyarsky:06c}
are derived from the observations of diffuse X-ray background, with the help
of non-imaging instruments, HEAO-I~\citep{Gruber:99}. The current status of
astrophysical observations in summarized in~\cite{Ruchayskiy:07}.

In this paper we use the spectrometer SPI on board of INTEGRAL satellite to
place restrictions on parameters of decaying DM in the mass range $40\kev -
14\mev$.  This range of masses is interesting, for example, the sterile
neutrinos, produced in the early Universe in the presence of large lepton
asymmetries~\citep{Shi:98} or through the inflaton
decay~\citep{Shaposhnikov:06}. It is also relevant for the case of gravitino
DM~\citep{Pagels:82,Bond:82}.

When the preparation of this paper was at its final stage,~\citet[hereafter \textbf{Y07}]{Yuksel:07} published their work, which used the results of~\citet[hereafter \textbf{T06}]{Teegarden:06}
to place restrictions on the parameters of sterile neutrino DM in the range
$40-700\keV$. We discuss it in more details in
Section~\ref{sec:discussion}.

\subsubsection*{SPI spectrometer}

The absence of the focusing optics significantly reduces the sensitivity of
the telescopes operating in the hard X-ray/soft \gr\ energy band. Most of the
instruments operating in this energy band use collimators and/or coded masks
to distinguish signals from the sources on the sky from the instrumental
background. Contrary to the focusing optics telescopes, both the source and
background signals are collected from the entire detector, which significantly
increases the irreducible background.

The focusing optics enables to significantly reduce the background only in the
studies of point sources. If the source under investigation occupies a large
fraction of the sky (e.g. the entire Milky Way galaxy), the performance of the
focusing and non-focusing instruments with the same detector collection area
are, in fact, comparable.

\begin{figure}
  \centering
  \includegraphics[width=\linewidth]{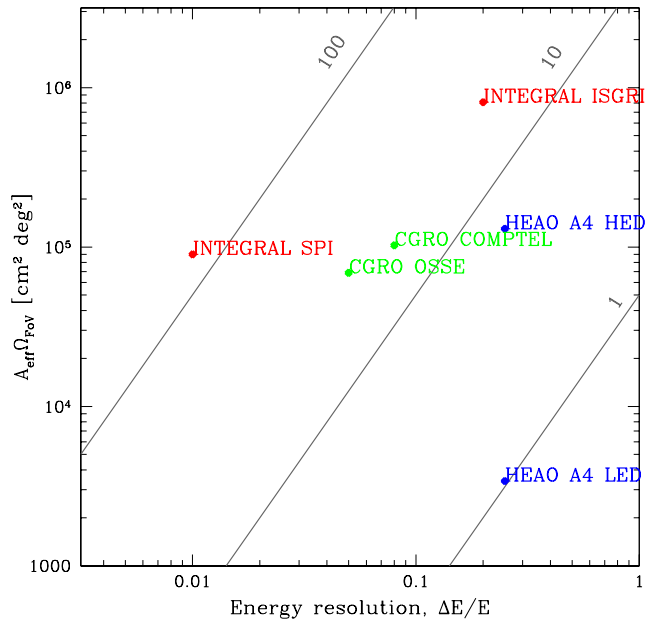} %
  \caption{Comparison of sensitivity towards the search of the narrow DM decay
    line for different instruments with the wide FoV. Diagonal straight lines
    show the improvement of sensitivity (by a factor, marked on the line) as
    compared with the HEAO-I A4 low energy detector (LED), taken as a
    reference.}
  \label{fig:spi_vs_heao}
\end{figure}

In the case of an extended source, emitting a narrow spectral line, an
efficient way of reduction of instrumental background is via the improvement
of the spectral resolution of the instrument (in the case of a broad continuum
background spectrum, the number of background counts at the energy of the line
is proportional to the spectral resolution $\Delta E$). The best possible
sensitivity is achieved when the spectral resolution reaches the intrinsic
width of the spectral line (see Fig.\ref{fig:spi_vs_heao} for the case of wide
FoV instruments and~\cite{Boyarsky:06f} for the case of narrow FoV
instruments).

In the case of the line produced by the DM decaying in the Milky Way halo, the
line width is determined by the Doppler broadening by the random motion of the
DM particles. The velocity dispersion of the DM motion in the halo is about
the rotation velocity of the Galactic disk, $v\sim 200$~km/s.  This means that
Doppler broadening of the DM decay line is about
\begin{equation}
  \label{eq:7}
\frac{\Delta E}{E}\sim \frac{v}{c}\simeq 10^{-3}\;.
\end{equation}
Thus, the optimal spectral resolution of an instrument searching for the DM
decay line produced by the Milky Way DM halo should be $\Delta E\simeq
10^{-3}E$.

Such optimal spectral resolution is almost achieved with the spectrometer SPI
on board of \intgr\ satellite, which has the maximal spectral resolving power
of $E/\Delta E\simeq 500$ and works in the energy range 20~keV --
8~MeV~\citep{spi}. SPI is a ``coded mask'' type instrument
with an array of 19
hexagonal shaped Ge detectors (of which only 17 are operating at the moment).

\begin{figure}
\begin{center}
  \includegraphics[width=\linewidth]{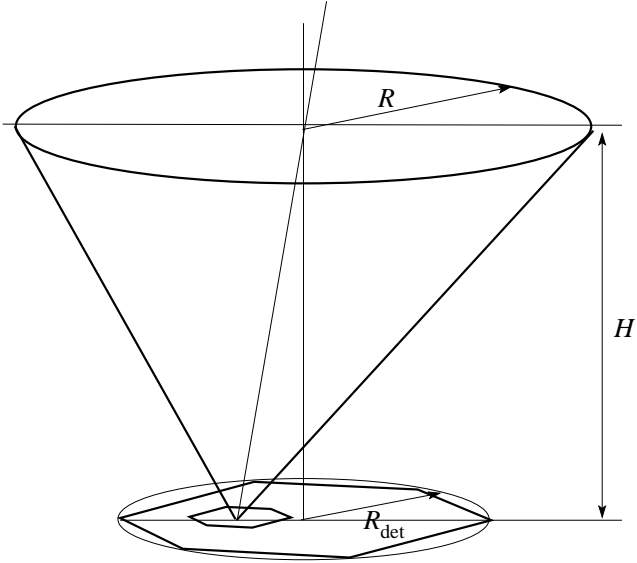}
\end{center}
\caption{The geometry of the SPI FoV.}
\label{fig:FoV}
\end{figure}
The SPI telescope consists of a coded mask inscribed into a circle of the
radius $R_{\rm mask}=39$~cm, placed at the height $H=171$~cm above the
detector plane and of the detector, which has the shape of a hexagon inscribed
into a circle of the radius $R_{\rm det}\simeq 15.3$~cm (see Fig.
\ref{fig:FoV}). The portion of the sky visible from each point of the SPI
detector (the so-called \emph{fully coded field of view}, FCFOV) has therefore
angular diameter
\begin{equation}
  \label{eq:12}
  \Theta_\fcfov=2\arctan\left[\frac{R_{\rm mask}-R_{\rm
        det}}H\right] \approx 16^\circ\;,
\end{equation}
while the portion of the sky visible by at least some of the detectors (the
\emph{partially coded field of view}, PCFOV) is
\begin{equation}
\label{eq:14}
  \Theta_\pcfov=2\arctan\left[\frac{R_{\rm mask}+R_{\rm
          det}}H\right] \approx 35^\circ\;.
\end{equation}
The solid angle spanned by the cone with this opening angle is
$\Omega_\pcfov=2\pi\Bigl(1-\cos(\Theta_\pcfov/2)\Bigr)\simeq 0.29$ (see
Fig.~\ref{fig:FoV}).
Wide field of view makes the SPI telescope suitable for the study of the very
extended sources, like the Milky Way DM halo.

\begin{figure}
\begin{center}
  \includegraphics[width=\linewidth]{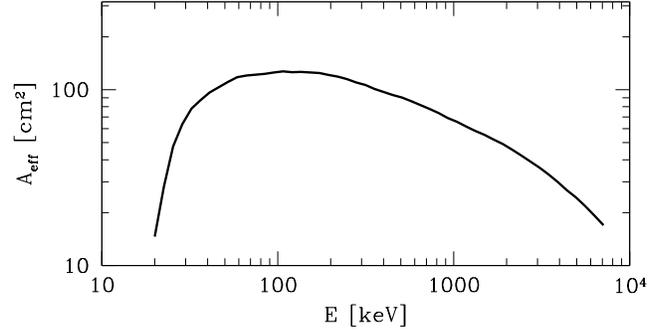}
\end{center}
\caption{The effective area of the SPI detector for an on-axis source, as a
function of the photon energy. The plot is produced by collective the
on-axis effective areas of the 17 SPI detectors from the instrumental
characteristics files.}
\label{fig:Aeffon}
\end{figure}

\section{The expected signal from the DM decay in the halo of
  the Milky Way.}
\label{sec:expected-signal}

The expected surface brightness of the DM decay line in a given direction on
the sky is a function of the angular distance $\phi$ between the given
direction on the sky and the direction towards the Galactic center (GC). It
can be calculated by taking the integral of the DM density profile
$\rho_\dm(r)$ along the line of sight (``\emph{column density}'')
\begin{equation}
  \label{eq:2}
  \CS_\dm(\phi) = \int\limits_0^\infty \hskip -.7ex dz \,
\rho_{\dm}\left(\sqrt{r_\odot^2-2zr_\odot\cos\phi+z^2}\right)\,,
\end{equation}
where $r_\odot\simeq 8.5\kpc$ is the distance from the Solar system to the GC
.  Angle $\phi$ is related to the galactic coordinates $(b,l)$ via
\begin{equation}
\label{eq:11}
\cos\phi =\cos b\,\cos l\;.
\end{equation}
Thus, the galactic center corresponds to $\phi=0^\circ$, the anti-center
$\phi=180^\circ$, and the direction perpendicular to the galactic plane to
$\phi=90^\circ$. The expected DM flux is given then by
\begin{equation}
\label{fdm}
\frac{dF_\dm\left(\phi\right)}{d\Omega}=
\frac{\Gamma_\dm E_\gamma}{4\pi\mdm} \CS_\dm(\phi)\;,
\end{equation}
where $\Gamma_{\dm}$ is the DM decay rate

In general, the surface brightness $F_{\rm DM}(\phi)$ is variable across the
telescope FoV.  This is especially true for a wide field of view (FoV)
instruments (like SPI).  In order to calculate the detector count rate, one
has to integrate flux~(\ref{fdm}) over the FoV and over the (effective)
detector area and then divide by the energy of the photons, $E_\gamma=\mdm/2$:
\begin{equation}
\label{R}
R=\iint\limits_\fov \hskip -.75ex d\alpha d\beta \,
\frac{A_\mathrm{eff}(E_\gamma|\alpha,\beta)}{E_\gamma}\frac{dF_{\dm}}{d\Omega}\bigl(\phi(\alpha,\beta)\bigr)\;,
\end{equation}
where $(\alpha,\beta)$ are the angular coordinates in the FoV, $A_\eff$ is the
effective area at energy $E_\gamma$ for the photons, coming from the direction
($\alpha,\beta$).

The effective area of the SPI detector (which is determined by the
transparency of the mask and the quantum efficiency of the detector) changes
with the photon energy. For an on-axis point source,
$$dF/d\Omega(\alpha,\beta)=f_0\delta(\alpha)\delta(\beta)\;,$$
the integral of
Eq. (\ref{R}) reduces to $f_0 A_{\rm eff,on}$, where $ A_{\rm eff,
  on}(E_\gamma)$ is the detector effective area \emph{for an on-axis source}.
Its dependence on energy $E_\gamma$ is shown on
Fig.~\ref{fig:Aeffon}.\footnote{The on-axis effective area is calculated by
  summing the energy-dependent on-axis effective areas of each of the 17
  operating detectors of SPI, extracted from the instrument's characteristics
  files.}

\begin{figure}
  \hskip -2.6cm\includegraphics[scale=.99]{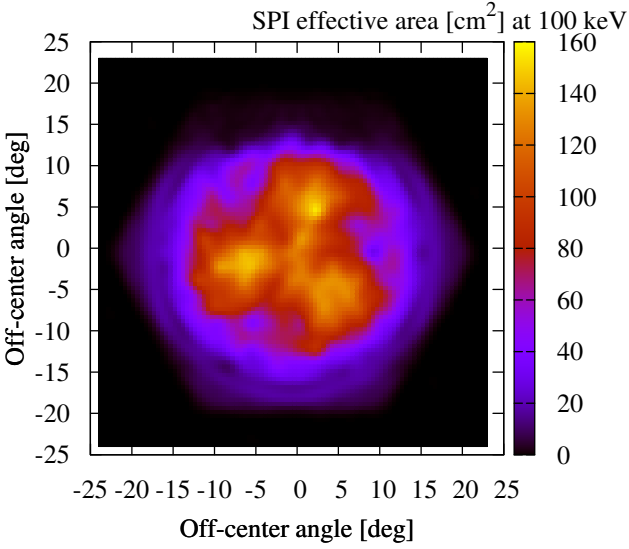} %
  \caption{Dependence of the effective area on the off-axis position of a
    (point) source. }
  \label{fig:area}
\end{figure}
In the general case of extended sources, evaluation of the detector count
rate~(\ref{R}) analytically is not possible because of the complicated
dependence of the effective area on the off-axis angle (shown on
Fig.~\ref{fig:area}). In the simplest case of an extended source with a
constant surface brightness $dF_{\dm}(\phi)/d\Omega=f_\text{ext}=const$, the
integral of Eq. (\ref{R}) reduces to the multiplication by the solid angle
$\Omega_{\pcfov}\simeq 0.29$ and the effective area, averaged over the FoV:
\begin{align}
  \label{eq:13}
  A_\text{eff,\,ext}(E_\gamma) &=\frac1{\Omega_\pcfov}\iint\limits_\fov \hskip
  -.75ex d\alpha
  d\beta \,A_\eff(E_\gamma|\alpha,\beta)\notag\\
  &\approx \kappa(E_\gamma) A_\text{eff,\,on}(E_\gamma) \;.
\end{align}
The numerical factor $\kappa(E_\gamma)$ depends on the energy and has to be
calculated via a numerical integration over the energy dependent off-axis response map of the SPI
detector. A reasonably accurate numerical approximation to $\kappa(E_\gamma)$ is
given by
\begin{equation}
  \label{eq:15}
  \kappa(E) \approx 0.165(E/\keV)^{0.11}\;.
\end{equation}
One can see that $\kappa\ll 1$ in all the energy interval. This is explained by
the fact that the detector area visible from a given direction on the sky strongly
decreases with the increase of the off-axis angle of this direction, so that the
sky-averaged effective area is much smaller than the on-axis effective area of the
detector.
%
Substituting (\ref{eq:15}), (\ref{eq:13}) into (\ref{R}) one finds that for an
extended source of constant surface brightness the detector count rate is
\begin{align}
\label{simple}
R_\text{ext} &= 2.73\times 10^{-5}\mathrm{\frac{cts}{s}}
\left[\frac{1\keV}{E_\gamma}\right]\\ 
&\times \left[\frac{A_\text{eff,\,ext}(E_\gamma)}{150\cm^2}\right]
\left[\frac{(dF_{\dm}/d\Omega)_\text{ext}}{10^{-15}\mathrm{erg/(cm^2\: s\:
      sr)}}\right] \nonumber
\end{align}

\subsection{Modeling the DM halo of the Galaxy}
\label{sec:mw}

The DM halo of the Galaxy has been extensively studied~\citep[see
e.g.][]{Kravtsov:98,Klypin:02,Battaglia:05}.  Various DM profiles, used to fit
observed velocity distributions, differ the most in the GC region.

It was shown in~\citet{Klypin:02,Battaglia:05} that the DM halo of the MW can
be described by the Navarro-Frenk-White (NFW) profile~\citep{Navarro:96}
\begin{equation}
  \label{eq:6}
  \rho_\nfw(r) = \frac{\rho_s r_s^3}{r(r+r_s)^2}\;,
\end{equation}
with parameters, given in Table~\ref{tab:nfw}.
The relation between virial parameters and $\rho_s$, $r_s$ can easily be found
(see e.g.  the Appendix~A of~\citealt{Boyarsky:06d}).

\begin{figure}
  \centering \includegraphics[width=\linewidth]{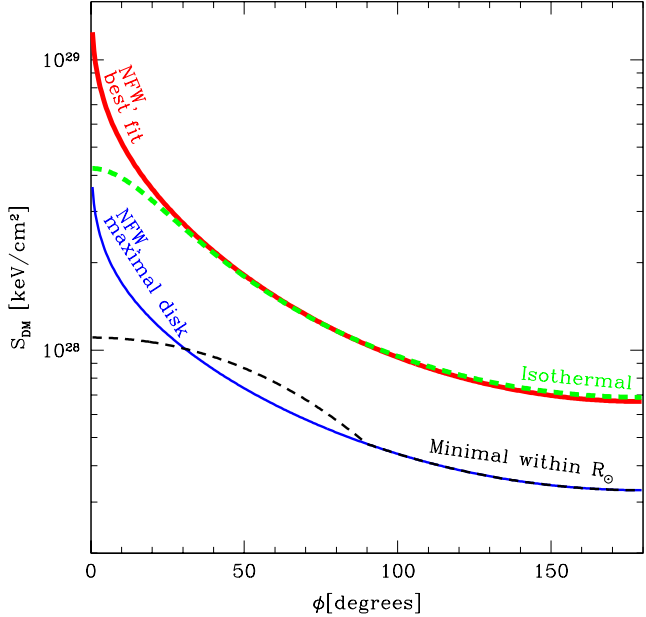}
    \caption{Expected column density  for various DM profiles: favored NFW
      profile (red thick solid line); NFW profile with the maximal disk (model
      $A_2$, see Table~\protect\ref{tab:nfw}) -- blue solid line; cored
      (isothermal) profile -- green thick dashed line; constant density within
      $r_\odot$ -- black dashed line). }
  \label{fig:profiles}
\end{figure}

To explore the uncertainty of the DM density profile in the inner part of the
Galaxy, we also describe the DM distribution in the MW via an isothermal
profile~\citep{Bahcall:80}:
\begin{equation}
  \label{eq:10}
  \rho_\iso(r) =\frac{v_h^2}{4\pi G_N}\frac1{r^2 + r_c^2} = \frac{\rho_0
    }{1+(r/r_c)^2}\;.
\end{equation}
The following parameters of isothermal profile reproduce the DM contribution
to the (outer parts of) Galaxy rotation curve $v_h = 170$~km/sec and $r_c =
4\kpc$~\citep{Boyarsky:06c,Boyarsky:06d} (i.e. $\rho_0 = 1.2\times 10^6
\frac{\kev}{\cm^3} \Bigl[\frac
{v_h}{170\;\mathrm{km/s}}\Bigr]^2\Bigl[\frac{4\kpc}{r_c}\Bigr]^2$).  These
parameters are consistent with those, from favored NFW models
of~\citet{Klypin:02,Battaglia:05}, i.e.  for $\phi\ge 90^\circ$ the difference
between isothermal model and NFW with preferred parameters was completely
negligible (less than 5\%) -- c.f.  FIG.\ref{fig:profiles}. Both types of
models provide the local DM density at the position of the Sun to be
$\rho_\dm(r_\odot) \simeq 0.22\gev/\mathrm{cm}^3$, which is close to the
existing
estimates~\citep{Kuijken:89a,Kuijken:89b,Kijken:89c,Kijken:91,Gilmore:98}.

The DM flux from a given direction $\phi$, measured by an observer on Earth
(distance $r_\odot\simeq 8.5\kpc$ from the GC), is given by
\begin{equation}
  \label{eq:19}
    \CS_\iso(\phi)\, {=}\, \frac {\rho_0 r_c^2}R\times \left \{
    \begin{array}{ll}
     \!\!\frac {\pi}{2} +
        \arctan\left(\frac{r_\odot\cos\phi}
        {R}\right) , &   \cos\phi \ge 0 \\
         \!\arctan\left(
  \frac{R}{r_\odot|\cos\phi|}\right), & \cos\phi < 0
        \end{array} \right. ,
\end{equation}
where $R=\sqrt{r_c^2 + r_\odot^2\sin^2\phi}$ and $\rho_0 r_c \simeq 1.5\times
10^{28}\kev/\cm^2$.

%
\begin{table*}
\caption{Best-fit parameters of NFW model of the MW DM halo. Max. disk model
  maximizes amount of baryonic matter in the inner $3\kpc$ of the MW halo
  ($M_\dm/(M_\text{disk}+M_\text{bulge}) = 0.4$ for the model $A_2$ and
  $M_\dm/(M_\text{disk}+M_\text{bulge}) = 0.14$ in the model $B_2$).}
\label{tab:nfw}
\centering
\begin{tabular}{c c c c c c}
\hline\hline
References & $M_\vir$ [$M_\odot$] & $r_\vir$ [kpc] & Concentration & $r_s$ [kpc] & $\rho_s$ [$M_\odot/\mathrm{kpc}^3$]\\
\hline
\cite{Klypin:02}, favored models ($A_1$ or $B_1$) & $1.0\times 10^{12}$ & 258 & 12 & 21.5 & $4.9\times 10^6$\\
\hline
\cite{Klypin:02}, Max. disk models $A_2$ & $0.71\times 10^{12}$ & 230 & 5 & 46 & $0.6\times 10^6$\\
\cite{Klypin:02}, Max. disk models $B_2$ & $0.71\times 10^{12}$ & 230 & 10 & 23 &$3.1\times 10^6$ \\
\cite{Battaglia:05}                               & $0.8^{+1.2}_{-0.2}\times 10^{12}$  & 255 &    18 & 14.2 & $11.2\times 10^6$\\
\hline
\end{tabular}
\end{table*}
%

The uncertainty of the DM radial density profile in the inner Galaxy stems
from the difficulty of separation between visible and DM contributions to the
inner Galaxy rotation curve.\footnote{When quoting results
  of~\cite{Klypin:02}, we do not take the effects of baryon compression on DM
  into account. While these effects make DM distribution in the core of the MW
  denser, any such computation is strongly model dependent.}  In order to get
the most conservative limit on the column density of the DM in the direction
of the GC, one can assume the following ``rigid lower bound'': while the DM
\emph{outside} the $r_\odot$ is described by the ``maximal disk'' model (model
$A_2$ of~\citealt{Klypin:02}), for $r\le r_\odot$ DM density remains constant
(so that the total DM mass within $r_\odot$ is the same as in the model $A_2$
of~\citealt{Klypin:02}). This gives
\begin{equation}
  \label{eq:1}
  \rho_\dm^\text{min} \simeq  3.9\times 10^6 \frac{M_\odot}{\kpc^3} = 0.146\times 10^6\frac{\kev}{\cm^3}\;.
\end{equation}
The surface brightness profile on the ``constant density'' model is shown in
black dashed line on the Fig.\ref{fig:profiles}. One can see that the
difference between the maximal ($\phi=0^\circ$) and the minimal
$\phi=180^\circ$) column densities is $\sim 3.4$ (as compared to $\sim 6$ for
isothermal model).  For comparison we show on Fig.~\ref{fig:profiles} expected
DM flux~(\ref{fdm}) for various profiles.
The minimal column density is of course the one in the direction of
anti-center: $\CS(\phi=180^\circ)\simeq 0.33\times 10^{28}\kev/\cm^2$. We see
that even for the minimal profile $S(\phi<30^\circ)\ge 10^{28}\kev/\cm^2$.

\subsection{DM decay line count rate}
\label{sec:count-rate-detector}

In the case of the Majorana sterile neutrinos of mass \mdm the DM decay
width is given by \citep{Pal:81,Barger:95}:\footnote{\label{fn:1}The quoted
  value of $\Gamma_\dm$ is for the Majorana sterile neutrino. In case of Dirac
  particle this value is 2 times smaller ~\citep[c.f.][]{Pal:81,Barger:95}.}
\begin{equation}
\label{Gamma}
\Gamma_{\dm}\simeq 1.3\times 10^{-32}\left[\frac{\sin^2
2\theta}{10^{-10}}\right]\left[\frac{\mdm}{1\kev}\right]^5\mbox{ s}^{-1}\ .
\end{equation}
Substituting (\ref{Gamma}) to (\ref{fdm}) we find
\begin{eqnarray}
\label{fdm1}
\frac{dF_{\dm}}{d\Omega}(\phi)&\simeq& 8.3\times 10^{-15} \mbox{ erg/(cm}^2\mbox{
  s sr)} \nonumber\\ \\
 &&\times
\left[\frac{\sin^22\theta}{10^{-10}}\right]
\left[\frac{\mdm}{1\mbox{ keV}}\right]^5
\left[\frac{\CS_\dm(\phi)}{10^{28}\mbox{ keV/cm}^2}\right]\nonumber
\end{eqnarray}
%
The lower bound on the DM decay line rate in SPI pointings toward the inner
Galaxy is calculated by substituting the column density
$\CS=10^{28}\kev/\cm^2$ (see Fig. \ref{fig:profiles}) into Eqs.
(\ref{simple}), (\ref{fdm1})
\begin{equation}
\label{rdm}
F_{\min}\simeq 3.0\times 10^{-6}\mathrm{\frac{cts}{cm^2\,s}}\left[\frac{\CS_\dm(\phi)}{10^{28}\mbox{ keV/cm}^2}\right]\left[\frac{\sin^22\theta}{10^{-10}}\right]
\left[\frac{\mdm}{1\mbox{ keV}}\right]^4
\end{equation}
The approximation of the constant surface brightness works well, if the
extended source has a core of the angular diameter exceeding the size of the
SPI partially coded FoV ($\Theta_{\rm PCFOV}\approx 17^\circ$ maximal off-axis
angle). Taking isothermal profile the angular size of the flat core of the
extended source is
\begin{equation}
\phi_{core}=\arctan(r_c/r_\odot)\simeq 25^\circ\;,
\end{equation}
which satisfies this constraint.

\section{Strategy of search for the DM decay line with SPI}
\label{sec:analysis}

The MW halo contribution to the DM decay signal represents the all-sky source.
Indeed, as the results of Section~\ref{sec:mw} show, the variability of the
signal over the sky may be as low as the factor $\sim 3$. This makes the
strategy of search of the DM decay signal different from any other types of
astrophysical sources: the point sources,
diffuse sources (e.g. $\sim 10^\circ$ Gaussian profile for $e^+e^-$
annihilation region,~\citet{Knodlseder:05}) or even the search for DM
annihilation signal~\citep[see
e.g.][]{Tasitsiomi:03,Boehm:03,Diemand:06,Sanchez:06,Carr:06}.

The problem gets exacerbated by the fact that during its motion, SPI is
irradiated by the charged high-energy particles (particles from Earth
radiation belt, Solar wind, cosmic TeV photons). As a result, the materials
(even detectors themselves) used for SPI construction start to radiate in
different energy regions (see subsection \ref{sec:spi-bkg}).
%
As a result any SPI spectrum consists of a broad continuum, which is a
combination of the sky and instrumental backgrounds, and of a set of the
instrumental background
lines~\citep{Attie:03,Diehl:03,Jean:03,Weidenspointner:03}. In order to detect
a spectral line produced by an astrophysical source one has to be able to
\emph{(a)} separate the continuum and line contributions to the spectrum and
\emph{(b)} separate the instrumental and sky signal contributions to the lines
found.


One can expect three \emph{a priori} situations:
\begin{enumerate}[\bf (I)]
\item DM decay line is strong (its \emph{equivalent width} much larger than
  the spectral resolution) and at its position there are no other strong lines
  (of either instrumental or astrophysical origin). Such a line, due to its
  presence in any SPI spectrum and its low variability over the sky can in
  principle be confused with some unknown instrumental line.

\item DM line is weak ($\sim 3-4\sigma$ detection over the continuum) but its
  position also does not coincide with any instrumental line.

\item DM decay line coincides with some instrumental line. To be able find
  such a line we need to model SPI instrumental background.
\end{enumerate}
To be able to work effectively with all these situations, we need to find the
way to separate the source and background contributions.


\subsection{Imaging}
\label{sec:imaging}

To distinguish source and background contribution to the signal, one often
uses imaging capabilities of an instrument.  If the size of a point or even an
extended source on the sky is smaller than the size of the SPI FoV one can (at
least, to some extent) use the imaging capabilities of the SPI instrument.  In
this case the coded mask, placed above the detector, partially screens the
individual detectors from the source, so that the source at a given position
on the sky produces different count rates in different detectors. One can find
the source flux by comparing the ratios of the actual count rates in different
modules of the detector to the ones predicted by the degree of screening of
the modules by the mask \citep[see][]{Dubath:05,Skinner:03}. It is a
challenge, however, to use the imaging capabilities of the SPI to separate the
astrophysical signal from the instrumental background if the size of the
extended source is comparable to the size of the SPI FoV~\citep[see e.g.][and
refs.  therein]{Knodlseder:05,Allain:06,Weidenspointner:07}.  Therefore for
our analysis we did not use any imaging capabilities of SPI, and to produce
spectra from some point in the sky we just collected all the photons, arriving
in the SPI FoV.

\subsection{SPI background modeling}
\label{sec:spi-bkg}

In the absence of imaging, the separation of the instrumental and astrophysical
contributions to the line spectrum requires some sort of background
modeling~\citep[see e.g.][]{Weidenspointner:03,Teegarden:04,Teegarden:06}.
Namely, for the background modeling we can use the fact that for any DM
distribution model the intensity of the DM decay line changes by a factor $\ge
3$ between the pointings towards the Galactic center ($\phi \sim 0^\circ$) and
anti-center ($\phi \sim 180^\circ$, see Sec.\ref{sec:mw}).  On the other hand,
if the line is of purely instrumental origin, there is no {\it a-priori}
reason why the strength of the line in the background spectra of the pointings
towards e.g.  the Galactic Anti-center should be different from the strength
of the line in the spectra of the pointings toward e.g the GC.  Thus, one
possible way to distinguish between the DM decay and instrumental origin of
the line is to study the variations of the line's strength depending on its
sky position (in the simplest case -- on the ``off-GC'' angle $\phi$, of the
pointing, Eq.~(\ref{eq:11})).

The situation becomes more complicated due to the fact that the instrumental
background (and thus the intensity of the instrumental lines) experiences
great variability in time (depending on the position in orbit, solar flares
and the solar activity period, degradation of the detectors, etc.,
c.f.~\citealt{Jean:03,Teegarden:04}).  As observations of different parts of
the sky can be significantly separated in time, one needs to use ``background
tracers'' to find the correct spatial dependence of the line
intensity~\citep{Jean:03,Teegarden:06}.  Without some sort of
``renormalization'' procedure, which corrects the absolute value of the line
flux using a measurement of a specific characteristics of the SPI instrument
as a ``calibrator'' of the flux, the $\phi$ dependence for any of the detected
lines contains no useful information.  There exist various ``background
tracers'' (Ge detectors saturation rates, anti-coincidence shield rates, rates
of certain background lines, see~\citet{Jean:03,Teegarden:04,Teegarden:06} and
refs. therein).

\subsection{Searching for the lines}
\label{sec:our-method}

To be able to detect strong DM line, which is not close in position to any
instrumental line (case~\textbf{I} above), we used the modification of the
method of background subtraction, described in TW06.  TW06 looked for $\gamma$-ray lines,
assuming different types of sources, from the point sources to the very diffuse sources
($10^\circ$ Gaussian, $30^\circ$ flat, etc.)  TW06 showed that the strong
background line at 198 keV can be used as a background tracer, if background observations are matched close in time to the corresponding ``source''
ones.  This allowed TW06 to cancel all strong instrumental lines with the
precision better than 1\%. TW06 detected \emph{no} emission line in such
background subtracted spectrum (apart from the 511~keV and 1809~keV) with the
significance above $3.5\sigma$.

We adopt the following modification of the TW06 method:
\begin{asparaitem}[--]

\item As the DM decay signal remains nearly constant within central
  $30-50^\circ$, the method of TW06, if applied directly, could cancel most of
  the DM signal.\footnote{For example for the most conservative DM
    distribution model, the difference of DM signals at $\phi=0^\circ$ and
    $\phi = 30^\circ$ is mere 8\%.}
  We therefore subtract the data (renormalized by the strength of 198~keV
  line) in the direction \emph{away} from the GC (off-GC angle
  $\phi>120^\circ$) from the ON-GC dataset (the angle $\phi \le 13^\circ$).

\item In the resulting ``ON--OFF'' spectrum we perform the search for the line
  with the significance higher than $3\sigma$.
\end{asparaitem}

This procedure allows to eliminate strong instrumental
lines with the precision better than few percents. At the same time any strong
DM line would remain in the ``ON--OFF'' spectrum. Indeed, even for the flattest
profile (Section~\ref{sec:mw}), the strength of the DM signal in the OFF
dataset is \emph{at least} 60\% weaker that of the ON dataset.  Therefore we
see, that the modification, described above, is indeed well suited for
searching of the strong DM decay line (case~\textbf{I}).

However, this method does not work well for the weak ($3-4\sigma$) lines, or
for the lines, whose position coincides with some instrumental line
(cases~\textbf{II-III} above). Indeed, in this case it is not possible to tell
whether the remaining line is the residual of the instrumental one or has the
astrophysical origin.  Below we will use an alternative method of analysis of
the detected lines, suitable for cases {\bf II} and {\bf III}.



\subsection{Analyzing a candidate line}
\label{sec:candidate-line}

Having detected a number of lines with the significance of $3\sigma$ and
above, we should decide which of them can be considered as ``DM decay line
candidates''. To this end we do the following.
\begin{asparaenum}[\bf a)]
\item We compare line flux for each of these lines with the flux of the same
  line in the ``ON'' spectrum. We decide that the line is a ``DM line
  candidate'' if the cancellation of the flux between ON and OFF datasets was
  worse than~10\%.\footnote{In principle, the DM line in ON--OFF spectrum
    should not cancel by more than $\sim 40\%$, while the background
    instrumental line should cancel better than 1\%. Thus the choice of the
    threshold to be around 10\% ensures that no DM decay line was thrown away
    while most of the instrumental lines disappeared.}
\item For any ``DM candidate line'' we construct its ``spatial profile'' (as
  described in details in the next Section) to check for the possibility of it
  to be a DM decay line (we also construct distribution of the line flux over
  the sky for all the unidentified lines from~\citealt{Weidenspointner:03}).
  Since the column density of the DM in the direction toward the GC should be
  higher than that of in the direction toward the Galactic anti-center, one
  should see a gradual decrease of the line strength with the increasing angle
  $\phi$.  We do not make any specific assumption about the DM density profile
  and do not try to fit the candidate line spatial profile to any particular
  model, but rather look if there is a general trend of decreasing
   intensity of the line with the increasing off-GC angle.
\end{asparaenum}



\section{Data reduction}
\label{sec:data-reduction}


\subsection{ON dataset}
\label{sec:dataset}

During its almost 5 years in orbit \intgr\ has intensively observed the inner
part of the Galaxy (Galactic Center, Galactic Bulge and the inner part of the
Galactic Plane) and collected about $T_{\rm exp}\sim 10$~Ms of exposure time
in the GC region. In our analysis of the inner Galaxy we used the publicly
available data (as of July 2007) from all \intgr\ pointings at which the angle
off the GC was at most $13^\circ$ and for which the SPI exposure time was
larger than 1~ksec.  This criteria selects 5355 pointings (or ``Science
Windows'', ScW), with total exposure time of 12.2~Ms, spread over the period
from February 2003, till April, 25, 2006. We call this dataset \texttt{"ON"}
dataset.

For each of the analyzed ScWs, we have extracted
photon (event) lists from {\tt spi-oper.fits} files 
and applied additional energy correction to convert the channel
number into photon energy, using \texttt{spi\_gain\_cor} tool from
standard Offline Analysis Software (OSA). We have binned the events
into narrow energy bins of the size $\Delta E_{\rm bin}=0.5$~keV to
generate the background counts spectra in each ScW, each revolution
and, subsequently, in the entire data set.


We then applied the ``sliding spectral window'' method to the line search (as
described e.g. in TW06 to produce a continuum subtracted
spectrum of the \texttt{"ON"} dataset.  Namely, at each given energy $E_0$,
one defines an energy interval $E_0-2\Delta E<E<E_0+2\Delta E$, where $\Delta
E$ is the SPI spectral resolution at a given energy, as a "line signal" energy
band.  For the (energy dependent) $\Delta E$ we used approximate formula from
SPI/INTEGRAL ground calibration of FWHM \citep{Attie:03}:
\begin{equation}
 \Delta E(E)=F_1+F_2\sqrt{E}+F_3E\label{eq:3}
\end{equation}
where $F_1=1.54$, $F_2=4.6\cdot 10^{-3}$, $F_3=6.0\cdot 10^{-4}$ and energy
$E$ is in keV. For $E=10^3$~keV $\mathrm{FWHM}\approx 2.3\kev$.

For each energy bin centered at an energy $E_0$ we have defined
the two adjacent energy intervals, $E_0-4\Delta
E<E<E_0-2\Delta E$ and $E_0+2\Delta E<E<E_0+4\Delta E$, and
postulated that the sum of the count rates in these two adjacent
energy bands gives the measure of the continuum count rate in the
energy band around $E_0$.  Subtracting the sum of the count rates in
the adjacent energy bands from the count rate in the "line signal"
energy band, we have calculated the continuum subtracted count rate at a given
energy $E_0$.  Doing such procedure at all the energies
$20~\mathrm{keV}<E_0<8~\mathrm{MeV}$, we have produced a ``continuum
subtracted'' SPI background spectrum.
In this spectrum we were able to identify most of the known instrumental
lines~\citep{Weidenspointner:03}.

\subsection{ON--OFF dataset}
\label{sec:on-minus-off}

\begin{figure}
  \centering %
  \includegraphics[width=\linewidth]{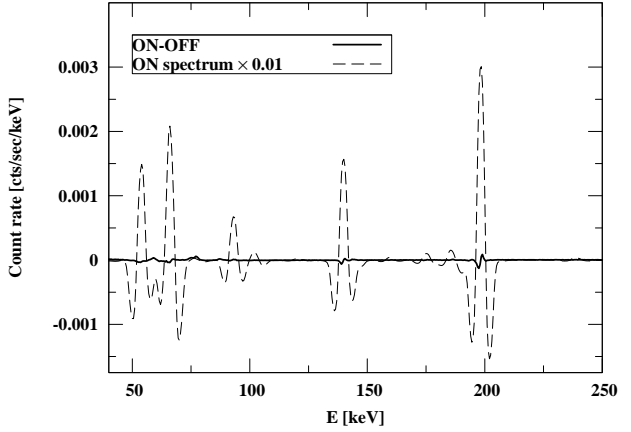} %
  \caption{Comparison of ON--OFF spectrum (thick solid line) with the
    $0.01\times {}$ the ON spectrum after the sliding window (thin dashed
    line). It can be seen that the instrumental lines are subtracted with the
    precision better than 1\%.}
  \label{fig:subtractionTW}
\end{figure}

Most of the lines found in the continuum subtracted background spectrum are of
the instrumental origin. To remove them, we matched each ScW in the ON dataset
with the pointing \emph{away} from the GC (galactic
coordinate $\phi>120^\circ$) -- OFF pointing.  %
As described by~TW06, the 198~keV line can serve as
good background tracer if the time duration between ON and OFF
observations is $\le 20$~days. We were able to match 3688 ON-OFF
pairs. For each ON--OFF pair we introduced normalizing coefficient $n$ for the OFF
spectrum in such a way that
 the strong instrumental line at 198 keV cancels
completely after subtraction of the OFF spectrum multiplied by the factor $n$
from the ON spectrum.  After that we
subtracted (renormalized) OFF ScW from the corresponding ScW from
the ON dataset.  This allowed us to remove most prominent
instrumental lines with the precision better than 1\% (c.f.
Fig.~\ref{fig:subtractionTW}).
To avoid contributions of strong astrophysical sources (such as e.g.
Crab) we threw out all pairs with negative total flux at 20--40~keV
range after subtraction.  Taking average over 2456  remaining
``good'' pairs we received the spectrum almost free from background
at energies above 200~keV. At low ($<200\kev$) energies we found
continuum component, which can be fitted with the simple power law:
\begin{equation}
  \label{eq:8}
  F(E) = F_0 \left[\frac{E}{100\kev}\right]^\alpha
\end{equation}
Parameters of this background were found to be
\begin{equation}
\label{eq:9}
\begin{aligned}
  F_0 &= (4.95\pm 0.05)\times 10^{-5} \mathrm{cts/s/cm^2}\\
  \alpha & = -(2.264 \pm 0.003)
\end{aligned}
\end{equation}
This continuum represents the residual contribution from all the set of the
astrophysical sources present in the Galactic Bulge.

\subsection{Systematic error}
\label{sec:systematic}
To estimate the systematic error of our ``ON-minus-OFF'' dataset, we computed
background around the ``tracer line'' of 198~keV.  We found that it does not
vanish. Thus, we estimated the systematic error as the error in the
normalization coefficient $n$ which would make the background zero within
systematic uncertainty.  This correction $\delta n$ can be found as follows.
Let $n$ be the coefficient, needed to cancel flux in the 198~keV line in ON
and OFF spectra:
\begin{equation}
  \label{eq:4}
  n = F_{ON}/ F_{OFF}
\end{equation}
where $F_{ON},F_{OFF}$ -- fluxes in 198keV line in ON and OFF ScWs
correspondingly.  The remaining non-zero $\delta F$ flux in the adjacent to
line position in ON-OFF spectrum, determines the uncertainty of the
coefficient:
\begin{equation}
  \label{eq:5}
  \delta n = \frac{\delta F}{F_{ON}}
\end{equation}
We found, that average value of $\langle\delta n\rangle$ is equal to
$\langle\delta n\rangle=1.1\cdot 10^{-3},\; \langle n\rangle \sim 1$. So, our
systematic error of final ON--OFF spectra at energy $E$ is $1.1\cdot
10^{-3}F_{OFF}(E)\approx 1.1\cdot 10^{-3}F_{ON}$. We add this systematic
uncertainty to the flux of ON--OFF spectrum in every energy bin.

\subsection{Obtaining $3\sigma$ restrictions}
\label{sec:obta-3sigma-restr}

At the energies at which no lines were detected (i.e. the ``continuum subtracted''
count rate did not deviate by more than $3\sigma$ from zero) we obtained the
$3\sigma$ upper limit on the possible flux from the DM decay. Above $\sim
200\kev$ the flux in the energy bin is zero within statistical errors,
therefore $3\sigma$ upper limit flux is given by statistical plus systematic
errors. Below $200\kev$ we put statistical restrictions above power law
continuum flux~(\ref{eq:8}), described in the Section~\ref{sec:on-minus-off}.
Using Eq.~(\ref{rdm}) one can derive the restriction on the sterile neutrino
mixing angle, implied by this upper limit. One should also take into account
that the subtraction of the OFF observations led to the reduction of the
expected DM signal. Taking the most conservative ``minimal'' model, described
in Section~\ref{sec:mw}, we see that the subtraction of the OFF signal leads
to about 40\% decrease of the expected DM signal.\footnote{To estimate this,
  we took the maximal column density for OFF observations at $\phi = 120^\circ
  - 17^\circ$.}
The resulting $3\sigma$ bound is shown on Fig.\ref{fig:on_limit}.

\subsection{Possible DM candidates}
\label{sec:lines}

\begin{table}
  \centering
  \begin{tabular}[c]{l|l}
    \hline
    Phase & Revolutions (start-stop)\\
    \hline
    1  & 042--092 \\
    \rowcolor[gray]{.7}
    & 096--140$^*$\\
   2 & 140-205\\
   \rowcolor[gray]{.7}
& 209-215$^*$ \\
   3   & 215-277\\
   4& 282-326\\
   5 & 330-395\\
   6& 400-446\\
   \hline
  \end{tabular}
  \caption{Splitting revolutions into phases in correspondence with  annealing
    phases and breakage of the detectors (revolutions, marked with the $^*$).}
  \label{tab:phases}
\end{table}

\begin{figure}
  \centering%
  \includegraphics[width=\linewidth]{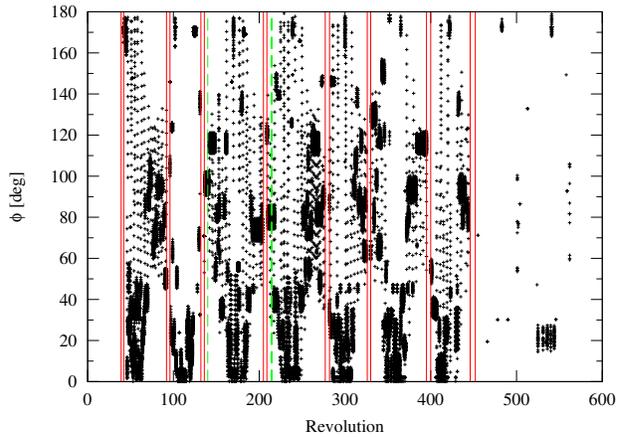} %
  \caption{Position on the sky as a function of revolution over 6 years of
    \intgr\ observations. The periods of annealing phases are shown in solid
    vertical lines. Two dashed lines indicate the revolution, during which 2ND
    and 17th of 19 SPI detectors have failed.}
  \label{fig:scw-over-sky}
\end{figure}

When analyzing ON--OFF spectrum, we found that almost all lines, present in ON
spectrum cancel with precision better than few percents.
We found 21 lines (see Table 3) that did not cancel by at least 90\%, (including known lines
at 511~keV and 1809~keV).  Apart from these 2 lines all other lines are
detected with low significance $3{-}4\sigma$.

As discussed in Sections~\ref{sec:our-method}--\ref{sec:candidate-line}, we
took all these lines as possible DM candidates and analyzed the dependence of
the line fluxes $F(\phi)$ on the off-GC angle $\phi$ of the pointing.  If the
DM distribution in the inner part of the Galaxy were known, it would be
possible to distinguish between the instrumental and DM decay origin of a line
by fitting $F(\phi)$ with a known profile calculated from the radial DM
density profile.  However, the details of the radial DM density profile in the
inner Galaxy are highly uncertain, and this prevents us from directly fitting
the model profile to the data.
We adopted a simple criterion which selects a DM decay candidate line: the
ratio of fluxes
\begin{equation}
  {\cal R}=\frac{F(0^\circ)}{F(180^\circ)}\ge {\cal R}_{min}\simeq 3.
\end{equation}
where ${\cal R}_{min}$ is the ratio of the DM decay line fluxes from the GC and
the Galactic anti-center in the ``minimal DM content'' model of DM
distribution.

\begin{figure}
  \centering%
  \includegraphics[width=\linewidth]{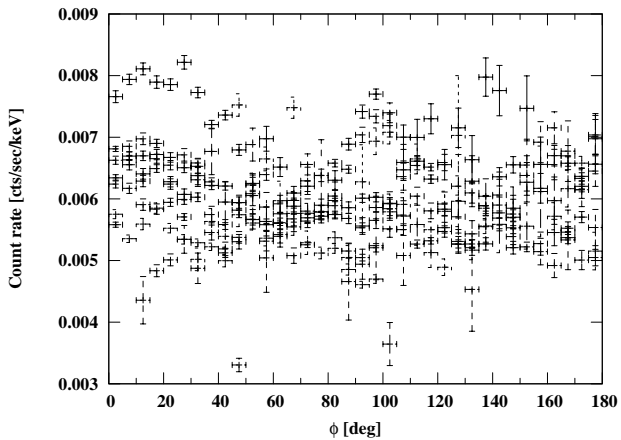} %
  \caption{Scatter of the flux data points for the line at $E=1068\kev$ as a
    function of the off-GC angle.}
  \label{fig:scatter}
\end{figure}

Since the observations at different off-GC angles are done during different time
periods, to properly study the dependence of the line flux on the off-GC
angle $\phi$ one should take into account the time variability of the response of
the SPI detectors. Several factors have to be taken into account. First,
 the SPI
instrument goes through a so-called ``annealing'' phase -- heating of the
detectors to recover from a radiative damage.\footnote{For details see SPI
  User Manual:\newline
  \href{http://isdc.unige.ch/Instrument/spi/doc/spi_um}{\text{http://isdc.unige.ch/Instrument/spi/doc/spi\_um}}.}
Next, two of the 19 SPI detector have ``died''.\footnote{Detector \# 2
  at revolution 140 and detector \#17 at revolutions 214-215.} %
The failed detectors also affect the response of their neighbors. To marginalize
the effects of the changing response of the SPI detector, we
split the entire data set into 7 periods, as shown on
Fig.~\ref{fig:scw-over-sky}. The intervals are summarized in the
Table~\ref{tab:phases}. As both detector failures occurred soon after the end
of an annealing phase, we chose to ignore revolutions 136 through 140 and 209
through 215. The period 096--140 does not cover the essential part of the sky
and therefore we skip it, leaving only 6 periods.

For each of the periods, shown in the Table~\ref{tab:phases}, we plot the
distribution of the line flux as a function of the off-GC angle $\phi$.  The
results are summarized on Fig.~\ref{fig:lines},~p.~\pageref{fig:lines}. One
can see that none of these lines exhibits clear trend of decreasing from
$\phi=0^\circ$ towards $\phi=180^\circ$.  For each line (and each phase) we
also compute the average flux $\bar F$, standard deviation $\sigma_F$ from the
average, minimum ($F_\min$) and maximum ($F_\max$). Our analysis shows that
\emph{(a)} $95-100\%$ of all points lie within 3$\sigma_F$ from the average
(thus, the data is consistent with having flat spatial profile) and \emph{(b)}
the scatter of the data ($F_\max - F_\min$) is much less than its mean value
$\langle F\rangle$.  Therefore, none of them cannot originate entirely from a
DM decay.  The corresponding numbers for each line and each phase are
summarized in Table~\ref{tab:profiles}, page~\pageref{tab:profiles}.

\begin{figure}
  \centering %
  \includegraphics[width=\linewidth]{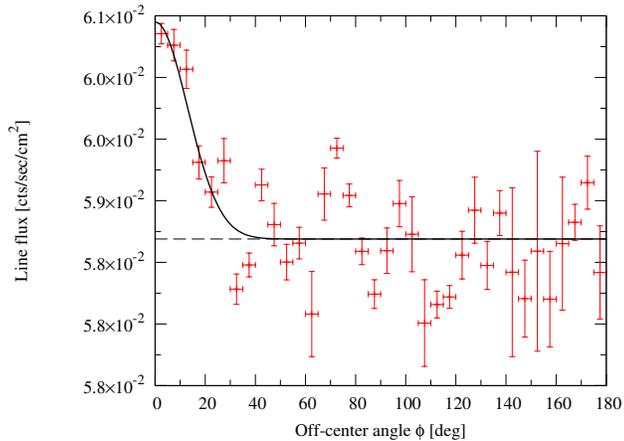} %
  \caption{Dependence of the intensity of the positron annihilation line at
    $E=511\kev$ on the off-GC angle. The solid line shows fit to the data in
    the form $\mathrm{const} + N e^{-\phi^2/(2\sigma^2)}$.}
  \label{fig:excess511}
\end{figure}

The positron annihilation line at $E=511\kev$ illustrates a situation, when a
line of astrophysical origin is superimposed on top of the strong instrumental
line. In this case, the data can be fitted by the constant, plus some
function, depending on assumed shape of the source. Fig.~\ref{fig:excess511}
shows the dependence of the flux of the 511 keV line on the off-GC angle of
the SPI pointing. One can see that for the pointings with the off-GC angle
less than $20^\circ$ (about the size of the PCFOV of SPI) the 511 keV line
flux contains a contribution from a sky source at the position of the GC,
while for the pointing at larger off-GC angles the astrophysical source is not
visible and the only contribution comes from the instrumental line, whose flux
does not depend on the off-GC angle of the pointing.


\begin{table*}
  \begin{tabular}[c]{r@{\extracolsep{-15pt}}l}
    \begin{minipage}{.5\textwidth}
      \begin{tabular}{|c|c|c|c|}
        \hline
        $E$ [keV] & Sign., $\sigma$ & $\Delta E$ [keV] & Identification\\
        \hline
        \input{line-table-tw120_v3.1.1.tab}
        \hline
      \end{tabular}
    \end{minipage}
   &
    \begin{minipage}{.5\textwidth}
      \begin{tabular}{|c|c|c|c|}
        \hline
        $E$ [keV] & Sign., $\sigma$ & $\Delta E$ [keV] & Identification\\
        \hline
        \input{line-table-tw120_v3.1.2.tab}
        \hline
      \end{tabular}
    \end{minipage}
  \end{tabular}
  \caption{Lines, detected in the ON--OFF spectrum with the significance $\ge
  3\sigma$. Lines, marked with $^\dagger$ cancel
  worse than by 90\% in the ON-OFF spectrum (as compared with their flux in ON
  spectrum) and thus represent a ``DM candidates''.
  The ``Identification'' column indicates the probable identification of the
  line in~\citet{Weidenspointner:03}.  Lines, marked with ``?'' are not
  identified in~\citet{Weidenspointner:03}, lines marked with ``??'' are not
  present in~\citet{Weidenspointner:03}.}
\label{tab:candidates}
\end{table*}

\begin{figure*}
  \centering \includegraphics[width=\textwidth]{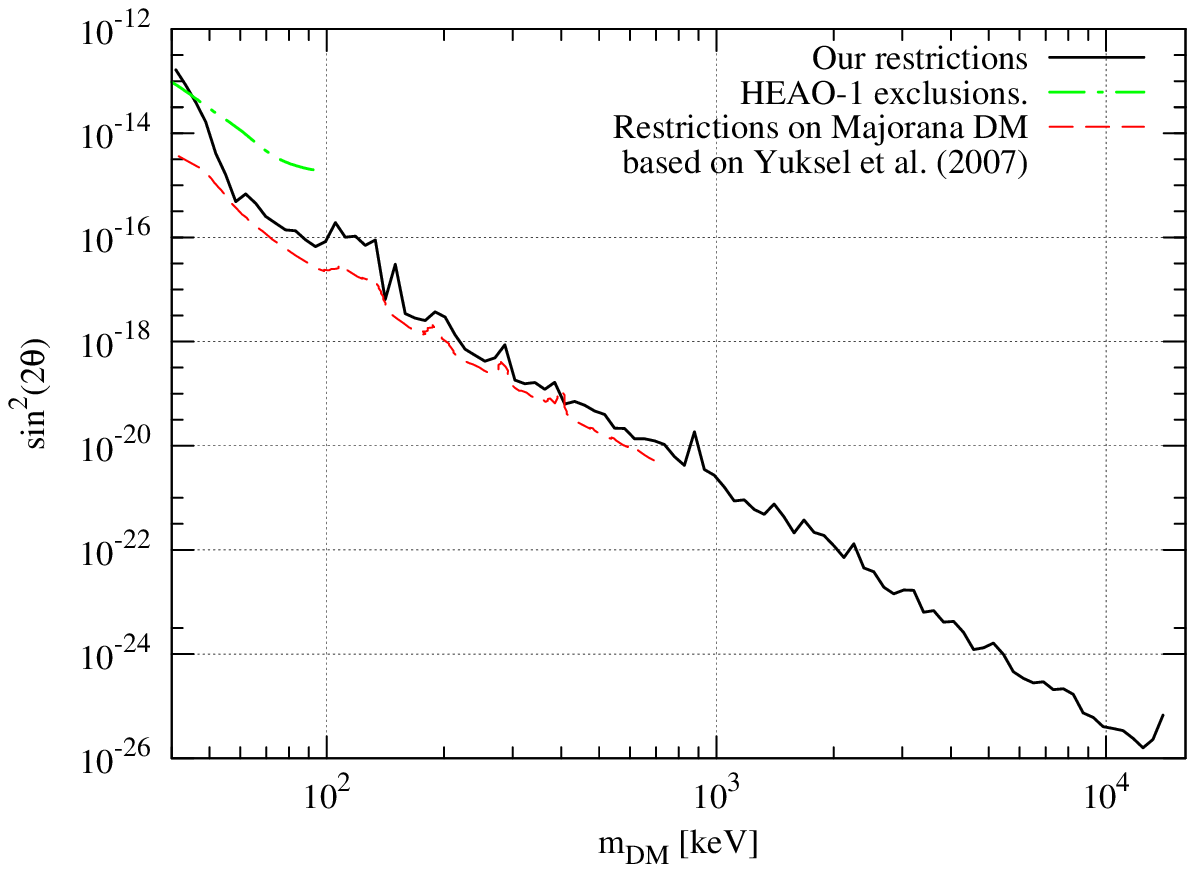}
\caption{Upper bound on the mixing angle of the DM sterile neutrino as a
  function of the sterile neutrino mass, obtained from the analysis of the
  background spectrum of the pointings toward the inner $13^\circ$ of the
  Galaxy. For masses $\le 700\kev$ the restrictions from~\citet{Yuksel:07}
  (divided by the factor 2, due to the Majorana nature of the DM) are also
  shown in dashed line (see Discussion). For masses $< 100\kev$ previous
  restrictions from HEAO-1~\citep{Boyarsky:06c} are also shown. The region
  \emph{above} the curve is excluded.}
\label{fig:on_limit}
\end{figure*}

\section{Results}
\label{sec:exclusions}

We analyzed the spectrum of SPI and found that none of the strong (i.e.
detected with significance above $5\sigma$) lines can be interpreted as that
of the decaying DM. This conclusion was based on the fact that variability of
these lines over the sky is less than 10\% (when moving from GC to the
anti-center, see Fig.\ref{fig:scatter}). At the same time for any realistic DM model such a variability
would be greater than at least 60\%.  Thus, we exclude the possibility that
one of the spectral lines, detected in the SPI background spectrum is a DM
decay line.

The non-detection of a DM decay line in the entire energy range of the SPI
detector has enabled us to put an upper limit on the parameters of the DM
particles. In particular, the $3\sigma$ upper bound on the mixing angle of the
sterile neutrino DM in the mass range 40~keV -- 7 MeV is shown on
Fig.~\ref{fig:on_limit}, p.~\pageref{fig:on_limit}.

\begin{figure}
  \centering%
  \includegraphics[width=\linewidth]{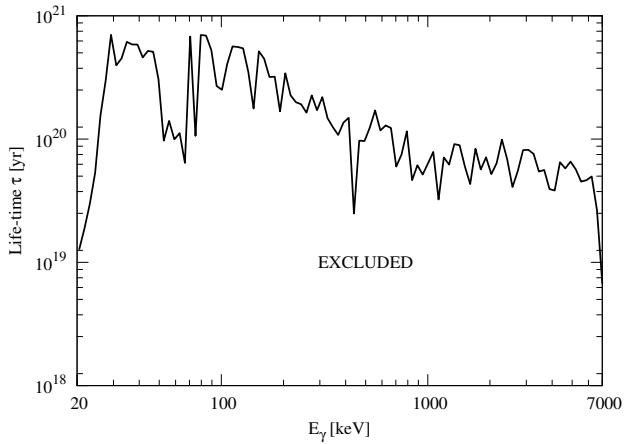}%
  \caption{Life-time of the radiatively decaying DM as a function of the
    emitted photon energy. Region \emph{below} the curve is excluded.}
  \label{fig:life-time}
\end{figure}

Our results are applicable to any decaying DM. To this end we also present the
restrictions on the DM life-time (with respect to the radiative decay) as a
function of the energy of emitted photon.  The corresponding exclusion plot is
shown on Fig.~\ref{fig:life-time}. For example, the gravitino can decay into
the neutrino and photon (similarly to the case of sterile neutrino) in
supersymmetric theories with broken R-parity. Such an interaction is generated
via the loop effects~\citep[see e.g.][]{Borgani:96,Lola:07}. The restrictions
on Fig.~\ref{fig:life-time} improve existing bounds on the life-time of such a
gravitino DM by several orders of magnitude (c.f.~\citealt{Borgani:96}).

\begin{figure}
  \centering %
  \includegraphics[width=\linewidth]{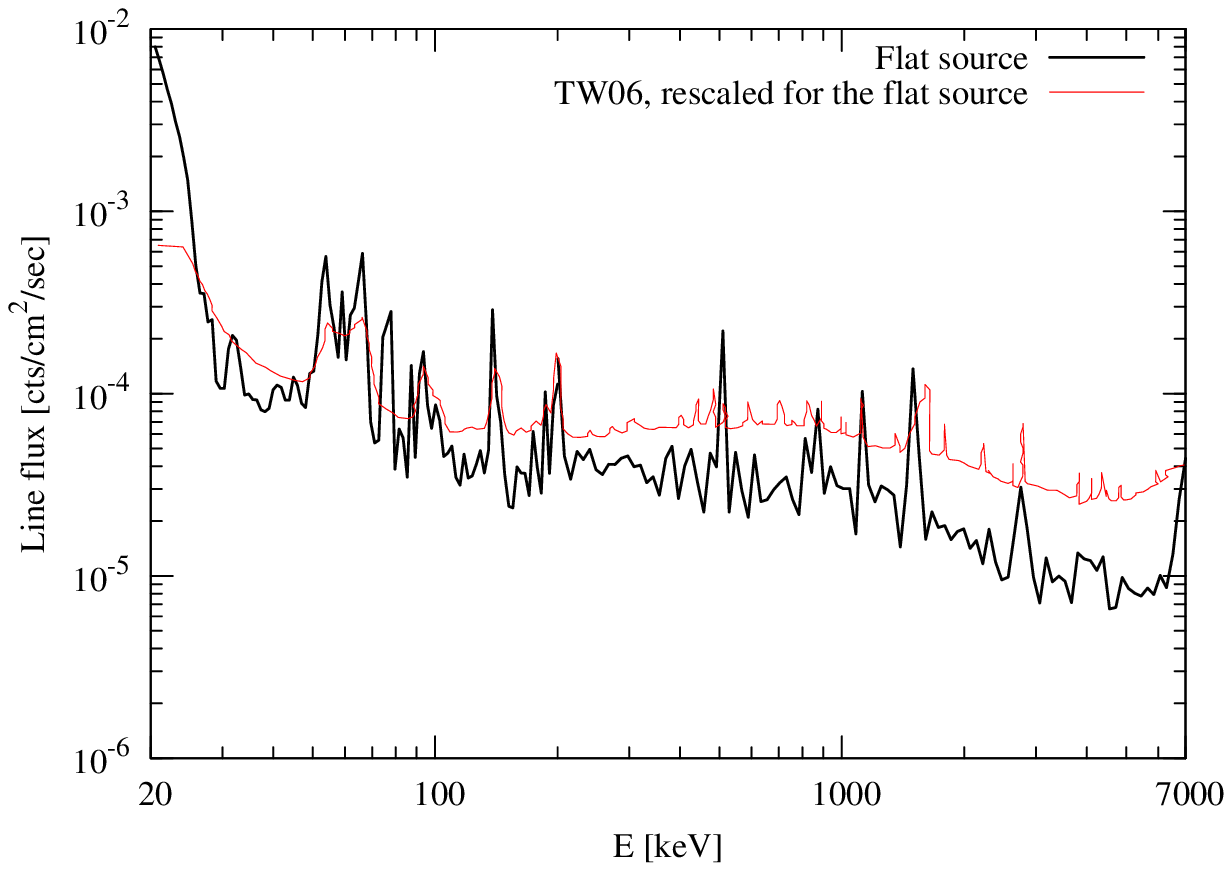} %
  \caption{$3\sigma$ sensitivity towards the line search in case of the flat
    diffuse source (thick black line). The results of
    \citet[Fig.~9]{Teegarden:06}, (rescaled to account for sensitivity towards
    the flat diffuse source, rather than $10^\circ$ Gaussian) are shown in
    thin red line.}
  \label{fig:sensitivity}
\end{figure}

To present our results in the form less dependent on a particular model of DM
distribution in the MW, we show the $3\sigma$ sensitivity towards the line
search on Fig.~\ref{fig:sensitivity}.  Note, that these results should be used
with care, as the sensitivity depends on the assumed spatial profile of the
source (because the effective area decreases with the off-axis angle, see
discussion in Section~\ref{sec:expected-signal}). The results, presented on
Fig.~\ref{fig:sensitivity} are valid for an extended source with the surface
brightness which varies on the angular scales larger than (or comparable to)
the size of the SPI field of view (black solid line).  This plot is analogous
to the Fig.~9 of \citet{Teegarden:06} (TW06).  However, a direct comparison of
the Fig. 9 of TW06 and Fig.~\ref{fig:sensitivity} is not possible, since TW06
have assumed a different morphology of the extended source ($10^\circ$
Gaussian). Explicitly taking into account the dependence of the effective area
of the SPI detector on the off-axis angle (see
Section~\ref{sec:expected-signal}), one can find that in order to make a
direct comparison between the two figures, one has to ``re-scale'' the results
of Fig.~9 of TW06 by an (energy dependent) factor of $\approx 1.5$. This
factor converts the sensitivity for the line, produced by a source with the
Gaussian surface brightness profile, into the one produced by a source of
approximately constant surface brightness (see red curve in
Fig.~\ref{fig:sensitivity}).

We have found a number of weak (with the significance $3{-}4\sigma$) lines in
the background-subtracted spectrum of SPI.  These lines cancel by worse than
90\% when subtracting OFF dataset (see Section~\ref{sec:data-reduction}).
Apart from it we have found in the background-subtracted spectrum two lines
with high significance -- known lines at 511~keV and at 1809~keV. Any of these
lines can in principle be a DM decay line.  We analyzed each of them, by
considering the profile of their intensity over the sky. Our analysis shows
that none of these lines could be \emph{pure} DM line (as their dependence on
the off-GC angle does not show any clear trend to decrease towards the
anti-center).  The possibility that some of these lines are the superposition
of instrumental and DM lines remains open.  Quantitative analysis of the
amount of DM flux admissible in a given line depends strongly on the model of
the DM distribution in the Milky Way halo. Therefore it was not conducted
here.

\section{Discussion}
\label{sec:discussion}

The purpose of this work was to understand how to search for the DM decay line
with the SPI spectrometer and to check that none of the strong lines, present
in the SPI background, was confused with the DM decay line. Our analysis shows
that all the strong lines were, indeed, of instrumental origin and provides
the upper bound on the flux of ``weak'' ($3{-}4\sigma$ above the background)
lines, which leads to the corresponding restrictions (see
Sec.~\ref{sec:exclusions}).  To further improve the results, one needs to work
with the weak lines (or lines, coinciding in position with instrumental ones).
To do this one needs more sophisticated procedures of subtraction of the
instrumental background (e.g. imaging). 

One of the most interesting cases of the coinciding instrumental and celestial
line is the positronium annihilation line at 511 keV. An excess of positron
annihilation emission on top of the strong instrumental line (related to
positrons annihilating inside the detector) was noticed long ago~\citep[for an
incomplete set of references see
e.g.][]{Prantzos:93,Milne:99,Cheng:97,Purcell:97,Knodlseder:05,Weidenspointner:06,Weidenspointner:07}.
There exist many attempts of explanation of this excess. In particular, it was
attributed to the annihilating or decaying DM~\citep[see
e.g.][]{Boehm:03,Hooper:03,Boehm:06,Frere:06,Picciotto:04,Rasera:05}.  The
sterile neutrino DM with the mass $m_s > 1\mev$ possesses decay channel
$N_s\to e^+e^-\nu$, with positrons annihilating either in flight or at rest,
by forming the positronium atom~\citep[see e.g.][]{Beacom:06,Sizun:06}. Thus,
it is possible that the decay of sterile neutrino DM contributes to such a
line.  The detailed analysis of this case will be reported separately.

It should be also mentioned, that the region of masses between $20\kev
\lesssim m_\dm \lesssim 40\kev$ remains inaccessible for the existing X-ray
missions. The strongest restrictions in this region were produced, using the
data of HEAO-1 mission \citep{Boyarsky:06c}.

When the work on this paper was at its final stage, the work of Y07 was published. Y07 obtained the restrictions on
parameters of sterile neutrino in the range 40 keV -- 700 keV. To facilitate
the comparison, we plot the restrictions of Y07 on Fig.~\ref{fig:on_limit},
(divided by the factor of 2 to translate them into the restrictions for the
Majorana, rather than Dirac sterile neutrino DM, see footnote~\ref{fn:1},
p~\pageref{fn:1}).  As the data, used in our work, has about 5 times longer
exposure than the \intgr\ first years data, on which the results of Y07 are
based, we could have expected  results stronger by a factor $\approx 2$ in our
case.  However, the Fig.~\ref{fig:on_limit} shows the opposite.  The reason
for this is as follows. For the SPI, the sensitivity towards the line search
from a particular source depends on the shape of the source. In particular,
the results of TW06, on which the work of Y07 was
based, were obtained under the assumption of a particular diffuse source
($10^\circ$ Gaussian). As any realistic DM profile is much flatter than the
$10^\circ$ Gaussian, the results of TW06 cannot be applied directly for the
case of the DM line search.  They should be rescaled to account for the
diffuse nature of the DM source~(c.f. Section~\ref{sec:exclusions}).
Apart from this, the estimated DM signal from the inner part of the Galaxy is
about 2 times stronger in Y07 than in our work.  As the DM signal in the
direction of the GC is the most uncertain, we have adopted the conservative
flat profile everywhere inside the solar radius, to minimize this uncertainty.

\subsection*{Acknowledgments}

We would like to thank B.~Teegarden and K.~Watanabe for useful discussion.
D.M. is grateful to the Scientific and Educational
Center\footnote{\url{http://sec.bitp.kiev.ua}} of the Bogolyubov Institute for
Theoretical Physics in Kiev, Ukraine, and especially to V.~Shadura, for
creating wonderful atmosphere for young Ukrainian scientists, and to Ukrainian
Virtual Roentgen and Gamma-Ray Observatory
VIRGO.UA\footnote{\url{http://virgo.bitp.kiev.ua}} and computing cluster of
Bogolyubov Institute for Theoretical
Physics\footnote{\url{http://grid.bitp.kiev.ua}}, for using their computing
resources. The work of D.M. was supported by the Swiss National Science
Foundation and the Swiss Agency for Development and Cooperation in the
framework of the programme SCOPES - Scientific co-operation between Eastern
Europe and Switzerland. The work of A.B. was (partially) supported by the EU
6th Framework Marie Curie Research and Training network "UniverseNet" (MRTN-
CT-2006-035863). O.R. would like to acknowledge support of the Swiss Science
Foundation.

\let\jnlstyle=\rm\def\jref#1{{\jnlstyle#1}}\def\aj{\jref{AJ}}
  \def\araa{\jref{ARA\&A}} \def\apj{\jref{ApJ}} \def\apjl{\jref{ApJ}}
  \def\apjs{\jref{ApJS}} \def\ao{\jref{Appl.~Opt.}} \def\apss{\jref{Ap\&SS}}
  \def\aap{\jref{A\&A}} \def\aapr{\jref{A\&A~Rev.}} \def\aaps{\jref{A\&AS}}
  \def\azh{\jref{AZh}} \def\baas{\jref{BAAS}} \def\jrasc{\jref{JRASC}}
  \def\memras{\jref{MmRAS}} \def\mnras{\jref{MNRAS}}
  \def\pra{\jref{Phys.~Rev.~A}} \def\prb{\jref{Phys.~Rev.~B}}
  \def\prc{\jref{Phys.~Rev.~C}} \def\prd{\jref{Phys.~Rev.~D}}
  \def\pre{\jref{Phys.~Rev.~E}} \def\prl{\jref{Phys.~Rev.~Lett.}}
  \def\pasp{\jref{PASP}} \def\pasj{\jref{PASJ}} \def\qjras{\jref{QJRAS}}
  \def\skytel{\jref{S\&T}} \def\solphys{\jref{Sol.~Phys.}}
  \def\sovast{\jref{Soviet~Ast.}} \def\ssr{\jref{Space~Sci.~Rev.}}
  \def\zap{\jref{ZAp}} \def\nat{\jref{Nature}} \def\iaucirc{\jref{IAU~Circ.}}
  \def\aplett{\jref{Astrophys.~Lett.}}
  \def\apspr{\jref{Astrophys.~Space~Phys.~Res.}}
  \def\bain{\jref{Bull.~Astron.~Inst.~Netherlands}}
  \def\fcp{\jref{Fund.~Cosmic~Phys.}} \def\gca{\jref{Geochim.~Cosmochim.~Acta}}
  \def\grl{\jref{Geophys.~Res.~Lett.}} \def\jcp{\jref{J.~Chem.~Phys.}}
  \def\jgr{\jref{J.~Geophys.~Res.}}
  \def\jqsrt{\jref{J.~Quant.~Spec.~Radiat.~Transf.}}
  \def\memsai{\jref{Mem.~Soc.~Astron.~Italiana}}
  \def\nphysa{\jref{Nucl.~Phys.~A}} \def\physrep{\jref{Phys.~Rep.}}
  \def\physscr{\jref{Phys.~Scr}} \def\planss{\jref{Planet.~Space~Sci.}}
  \def\procspie{\jref{Proc.~SPIE}} \let\astap=\aap \let\apjlett=\apjl
  \let\apjsupp=\apjs \let\applopt=\ao

\newcommand{\threeLines}[3]{%
  \includegraphics[width=.285\textwidth]{line#1}&
  \includegraphics[width=.285\textwidth]{line#2}&
  \includegraphics[width=.285\textwidth]{line#3}\\
} %
\begin{figure*}
  \centering
  \begin{tabular}[c]{ccc}
    \threeLines{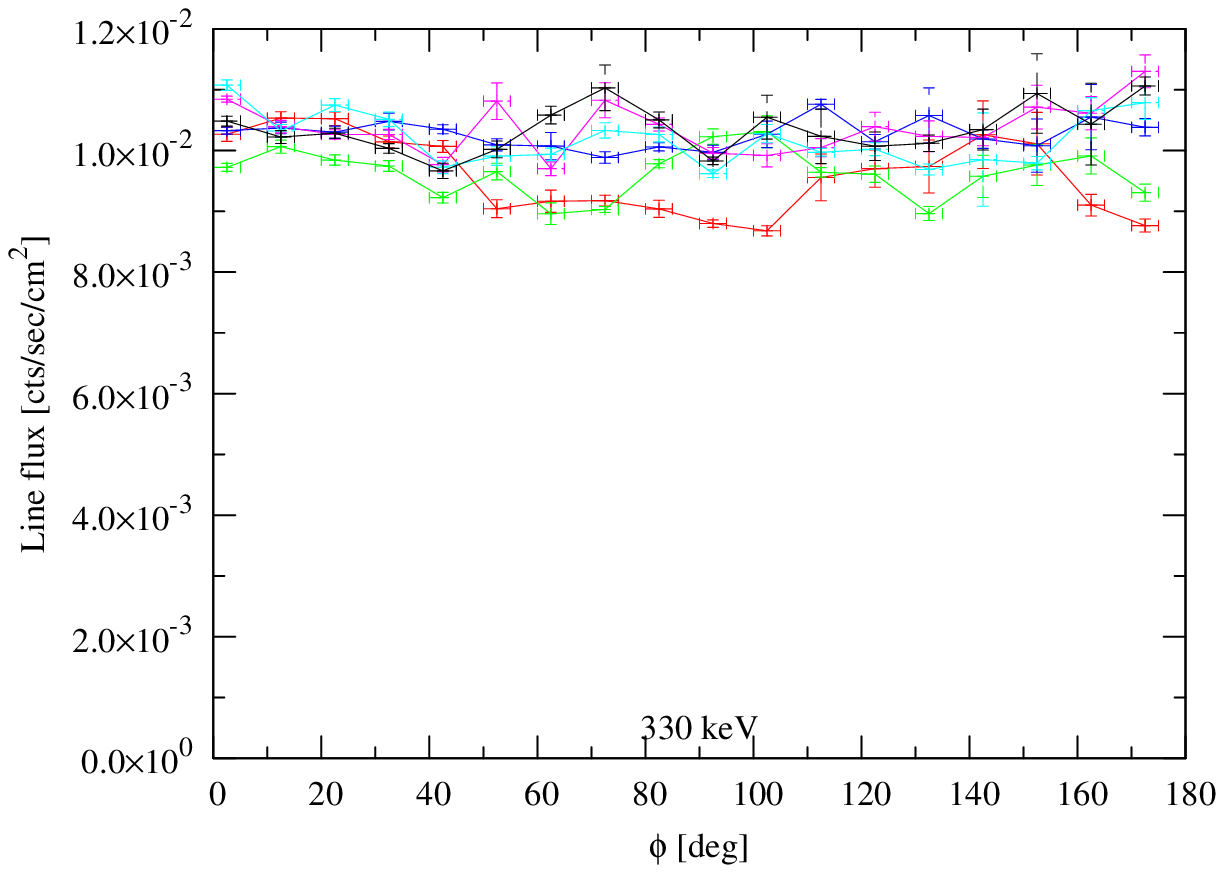}{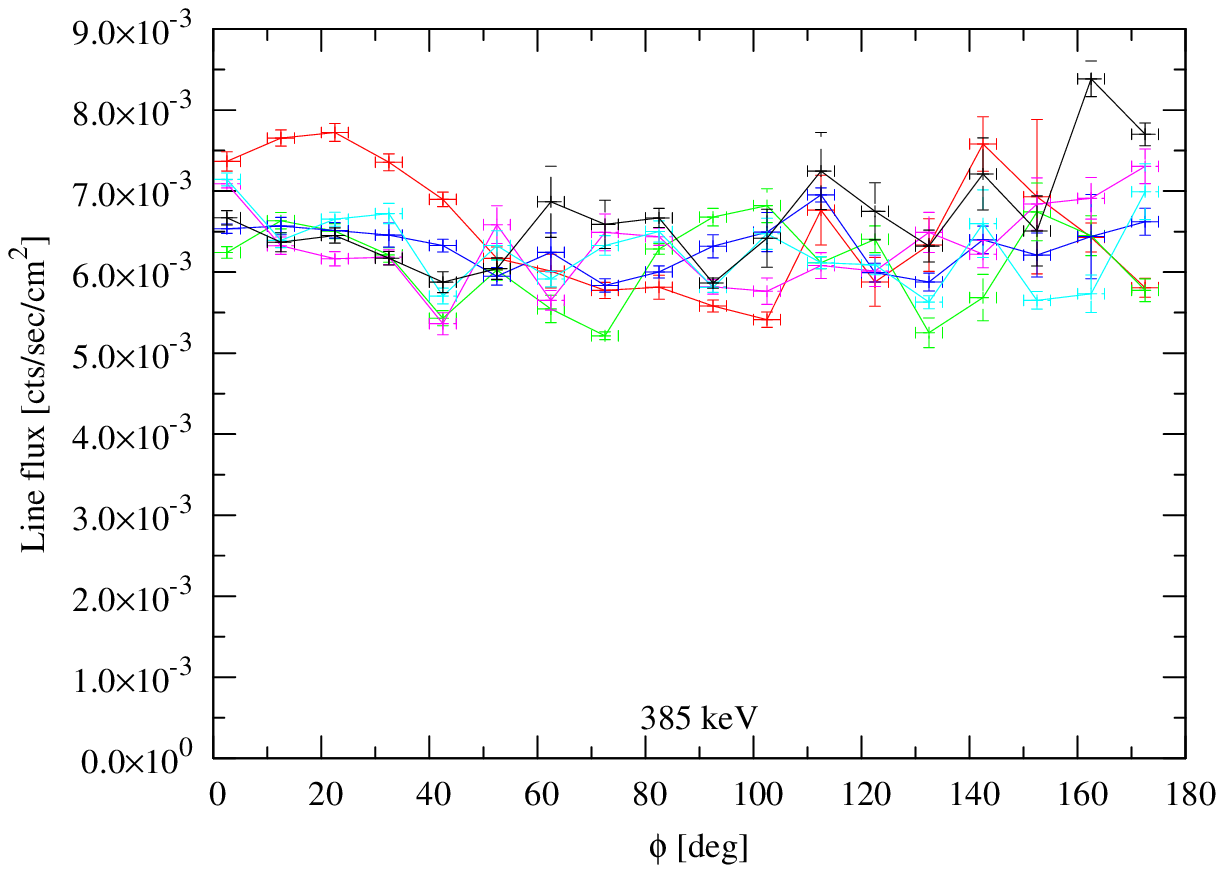}{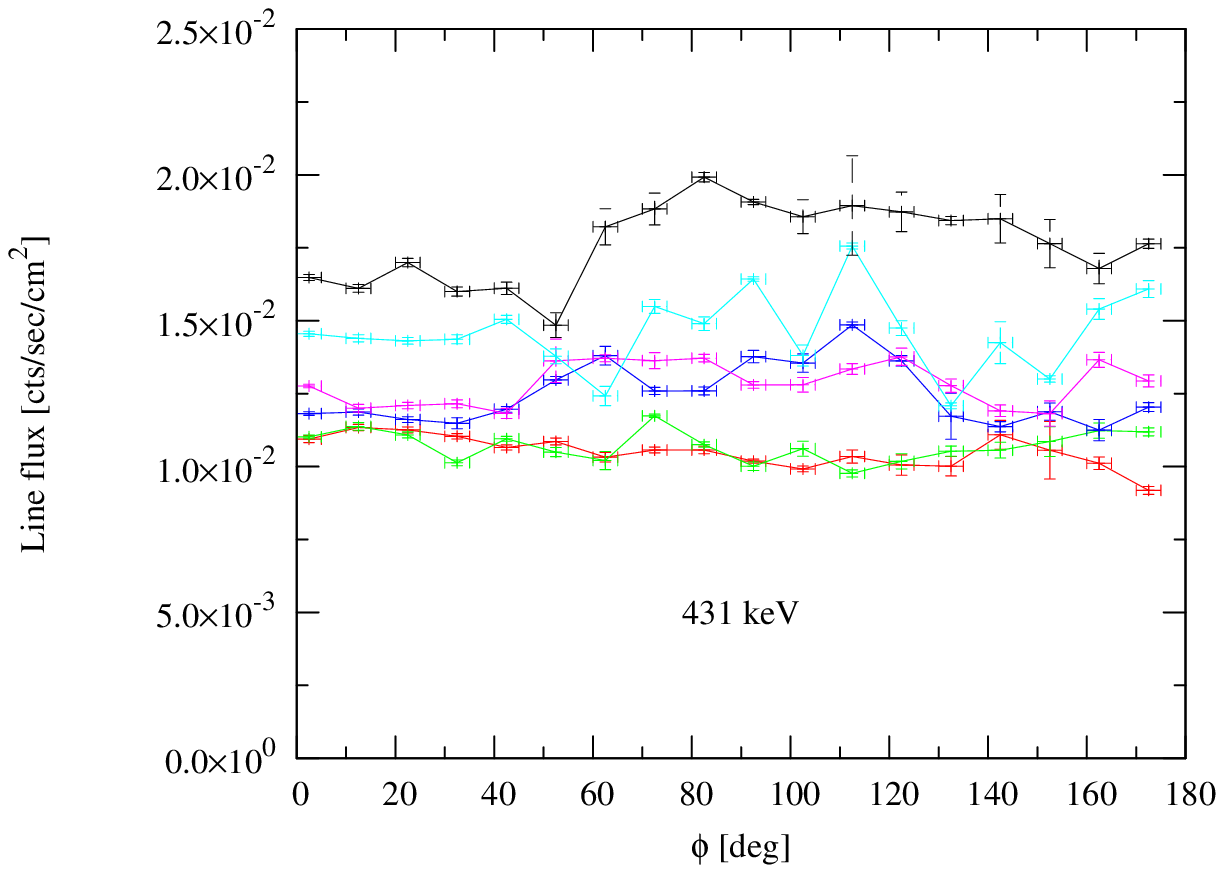}
    \threeLines{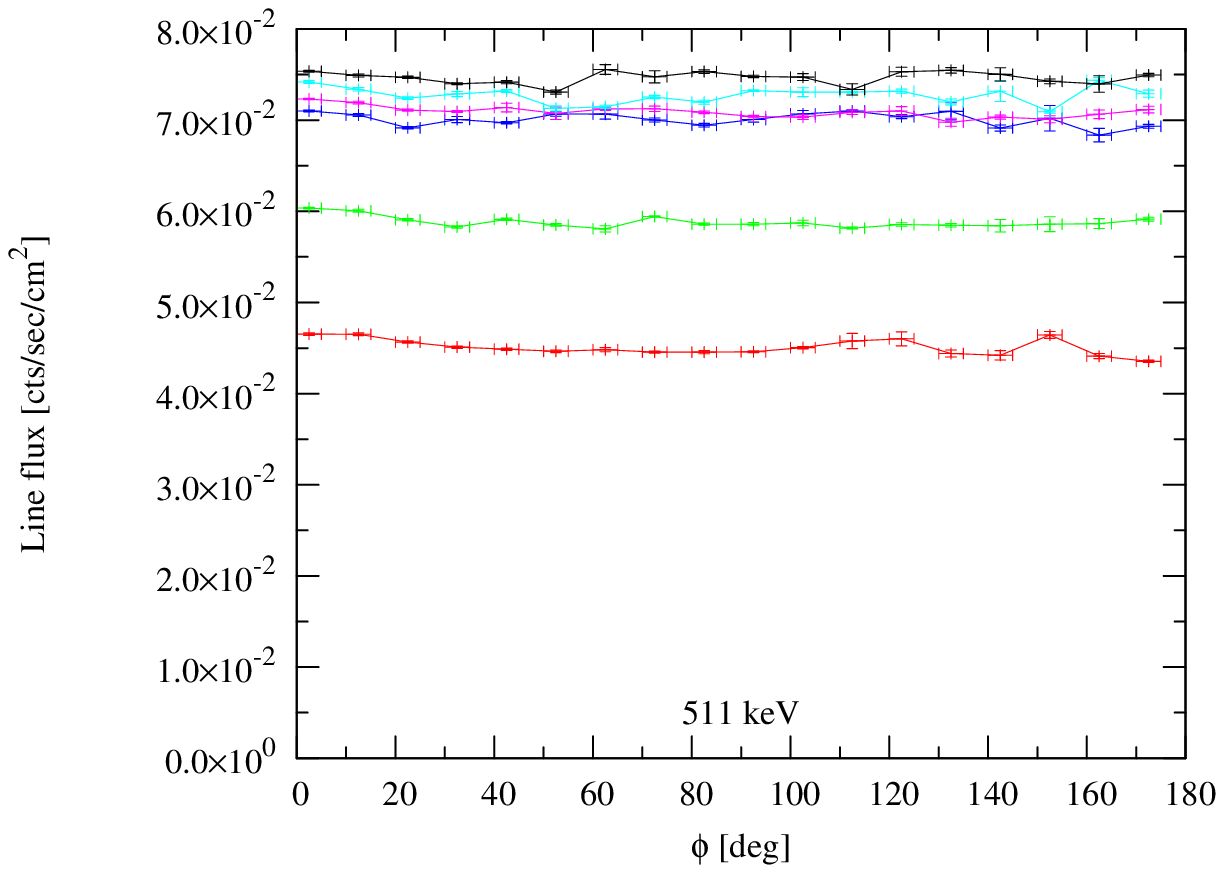}{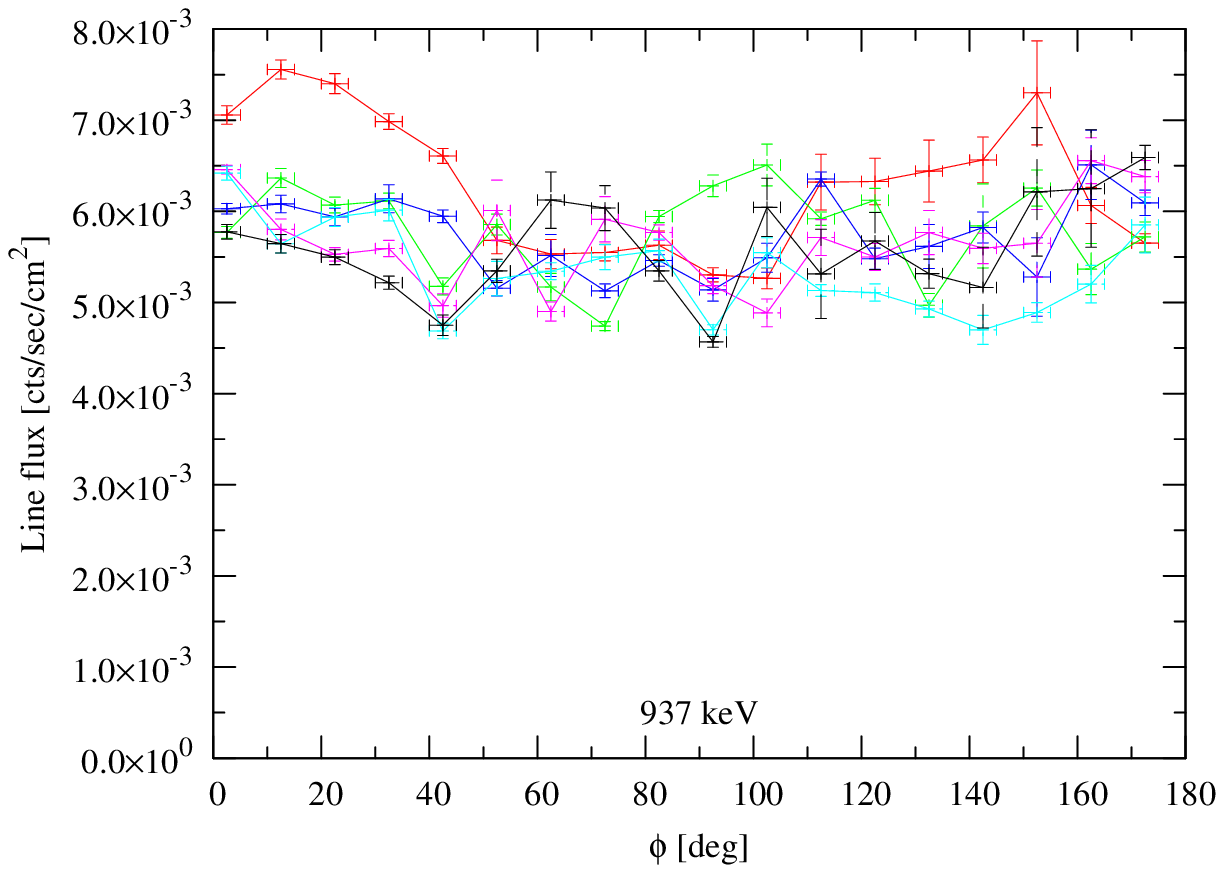}{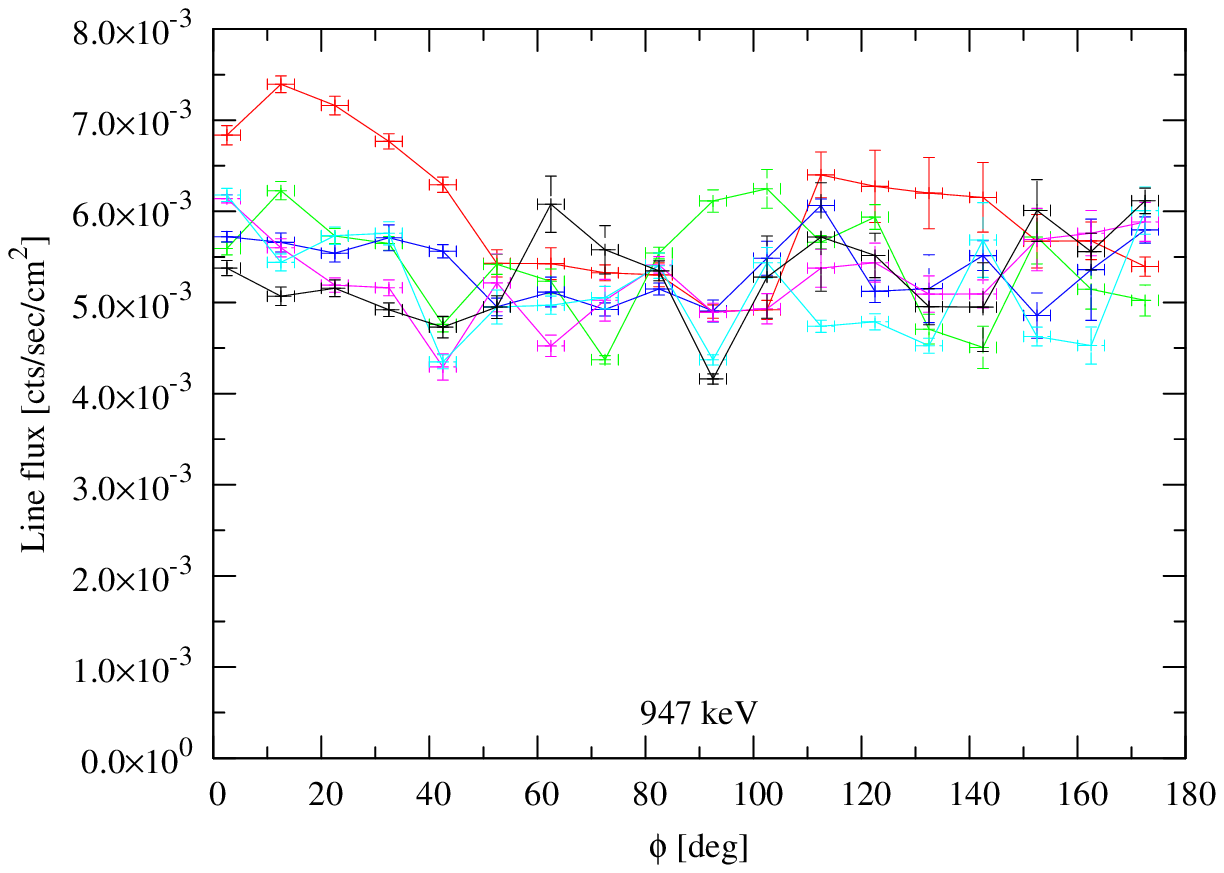}
    \threeLines{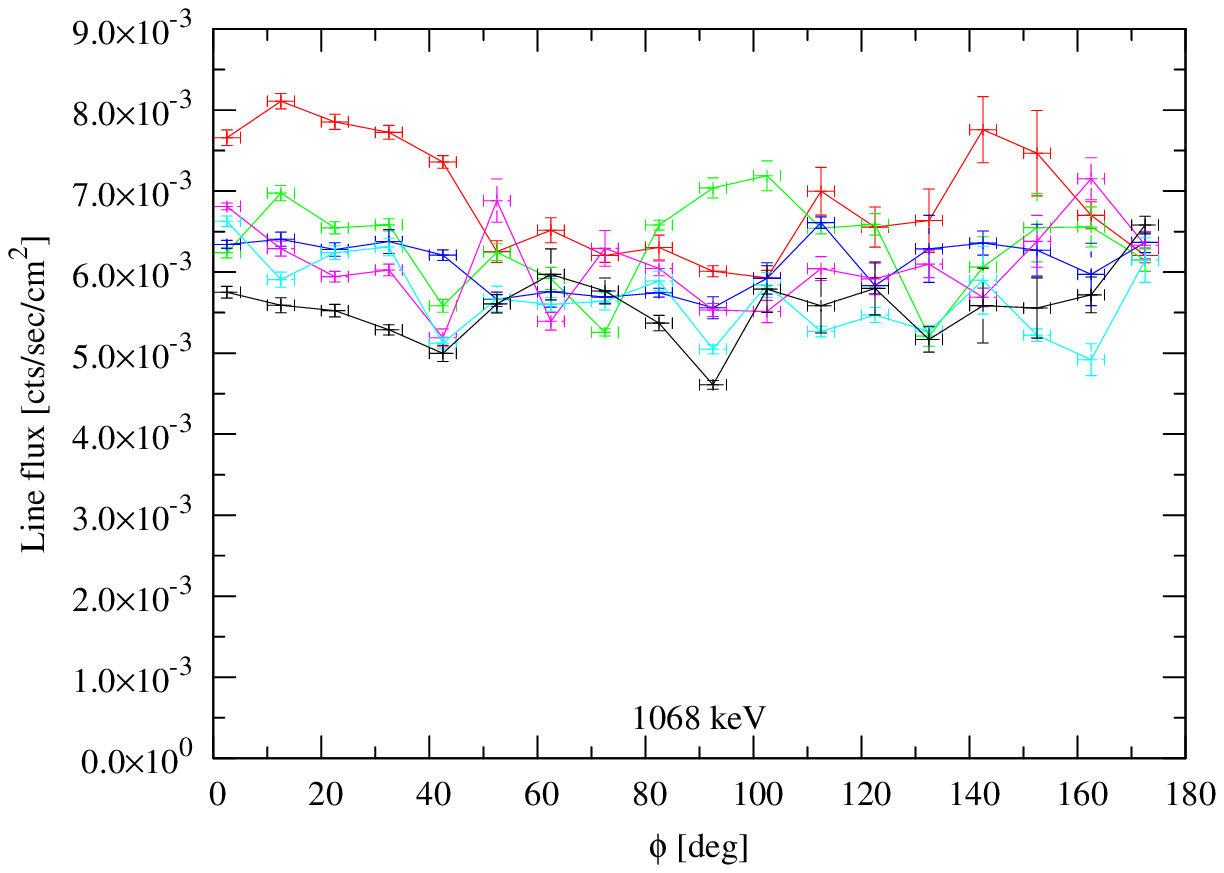}{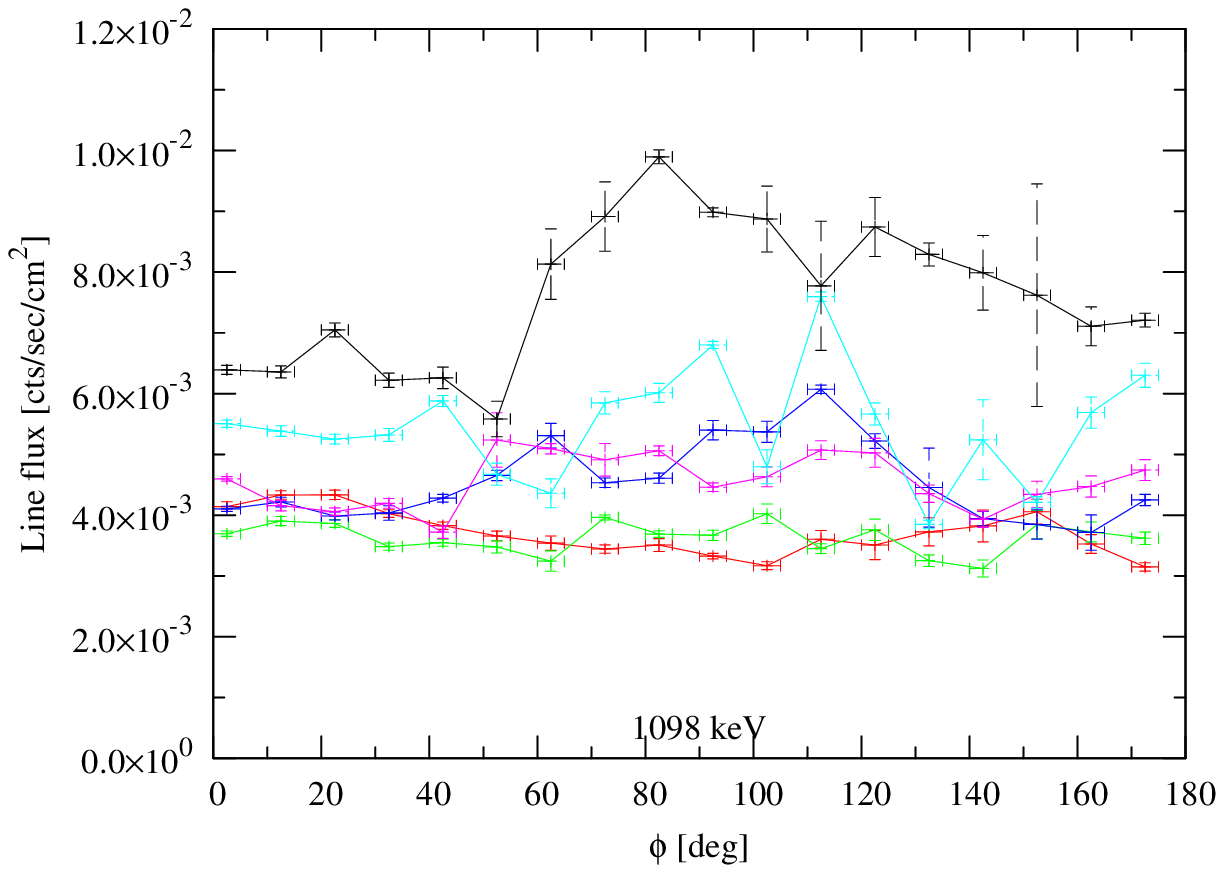}{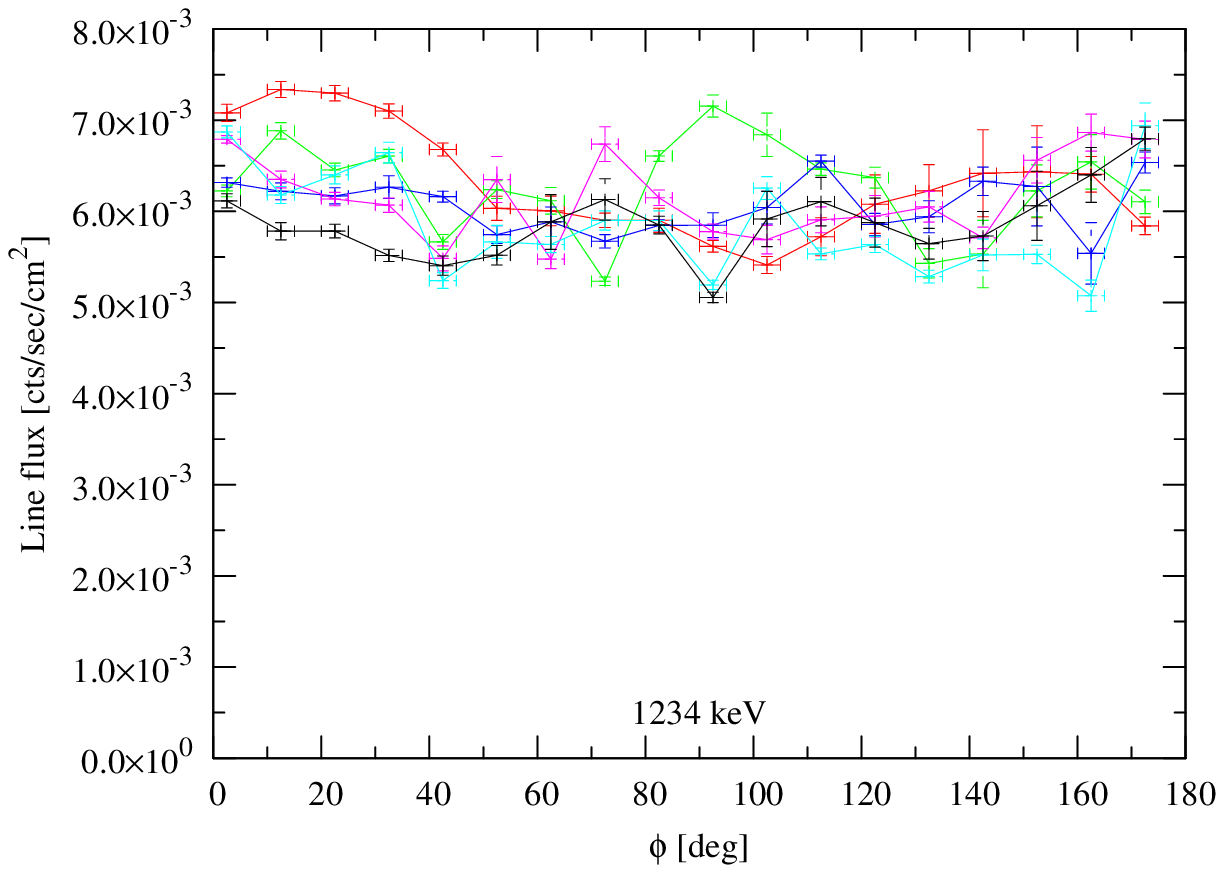}
    \threeLines{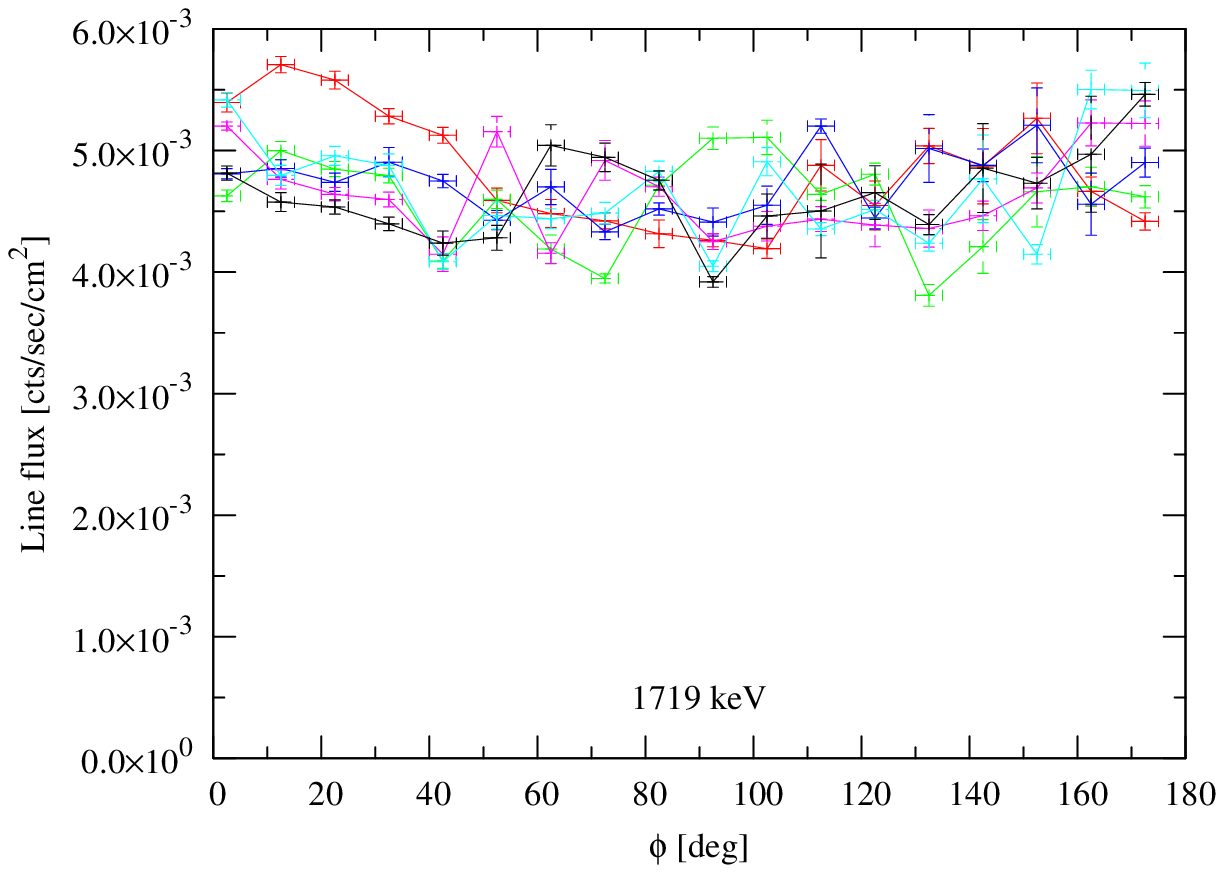}{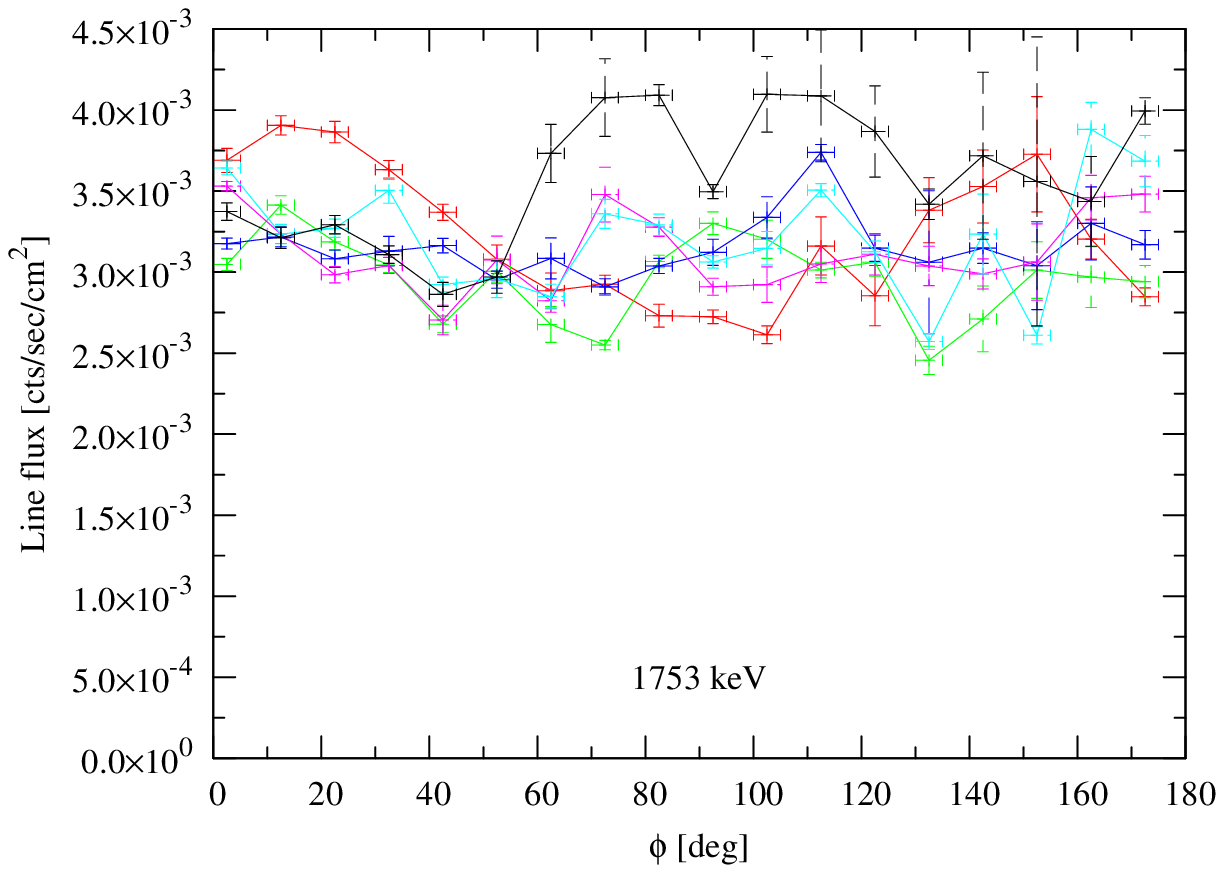}{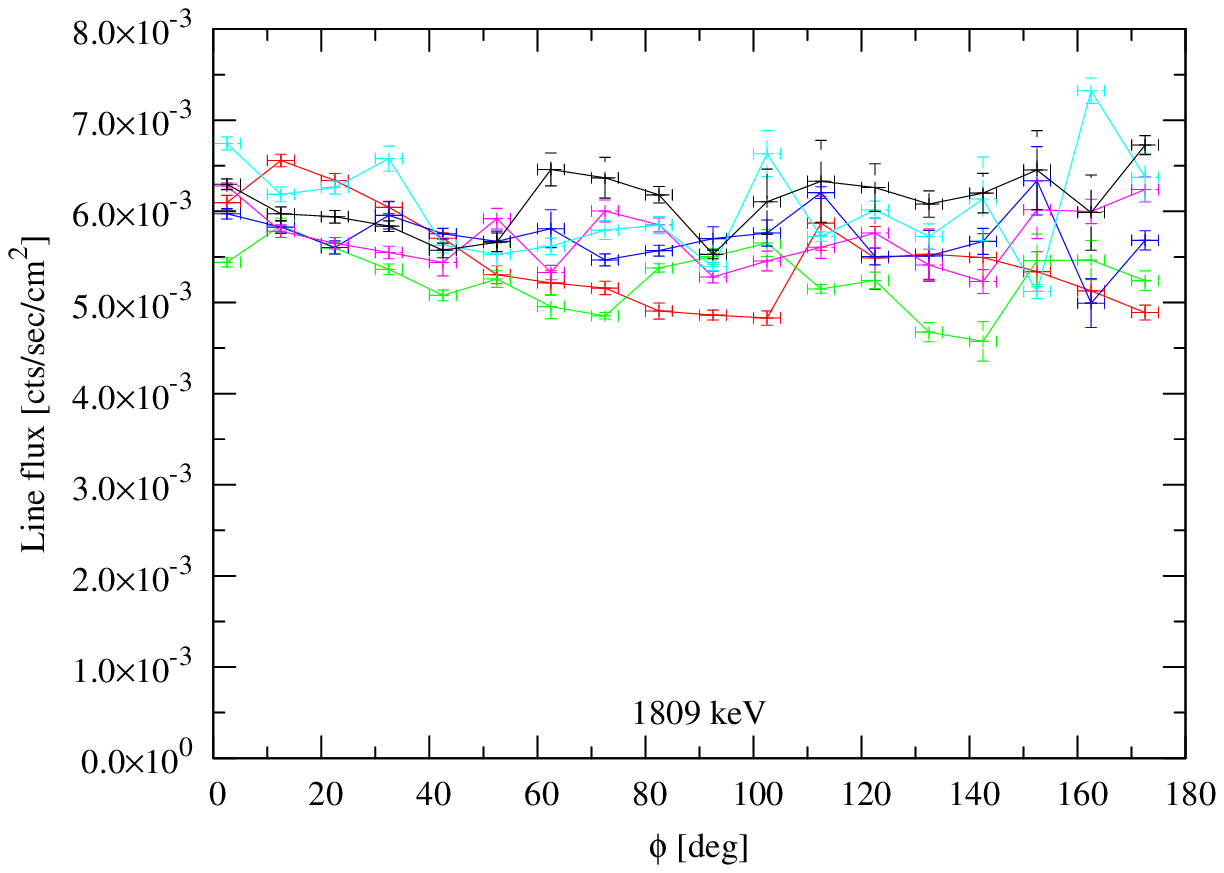}
    \threeLines{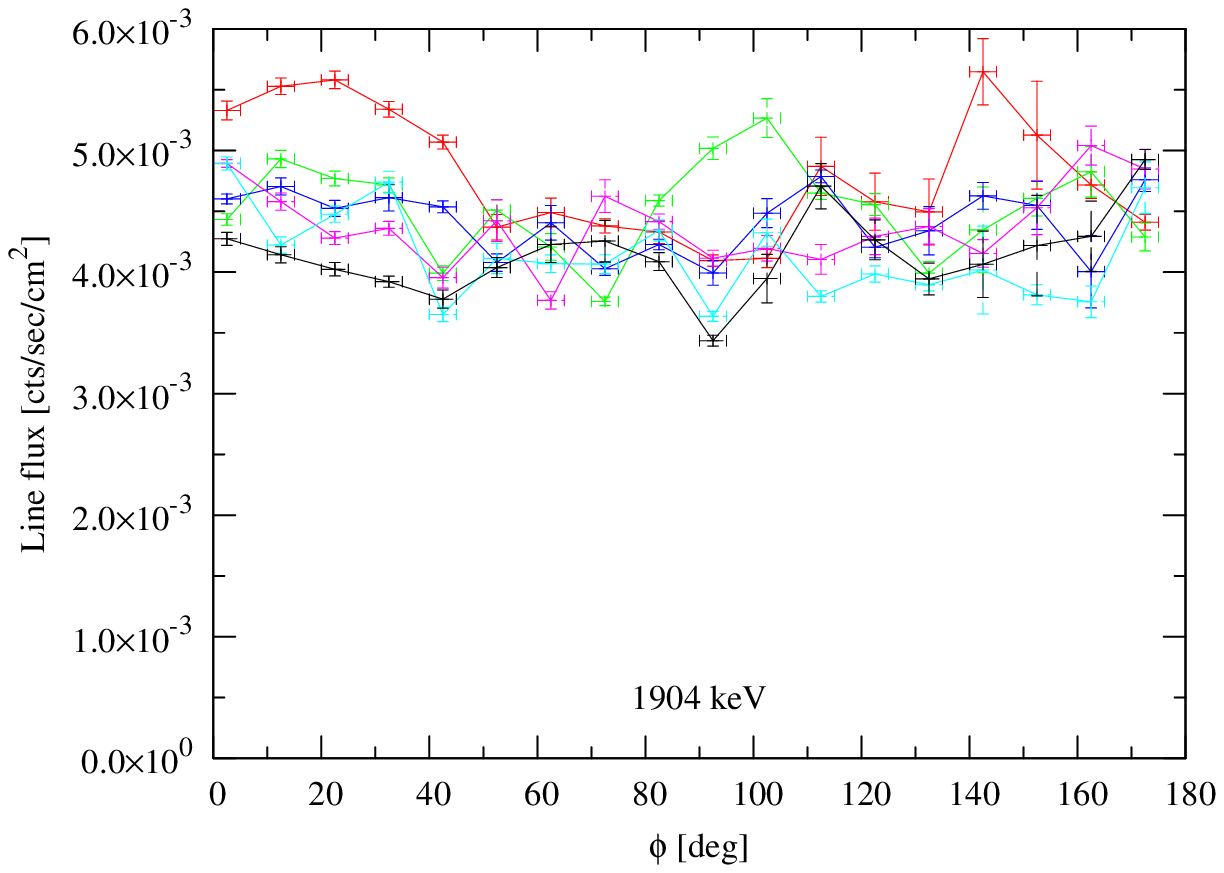}{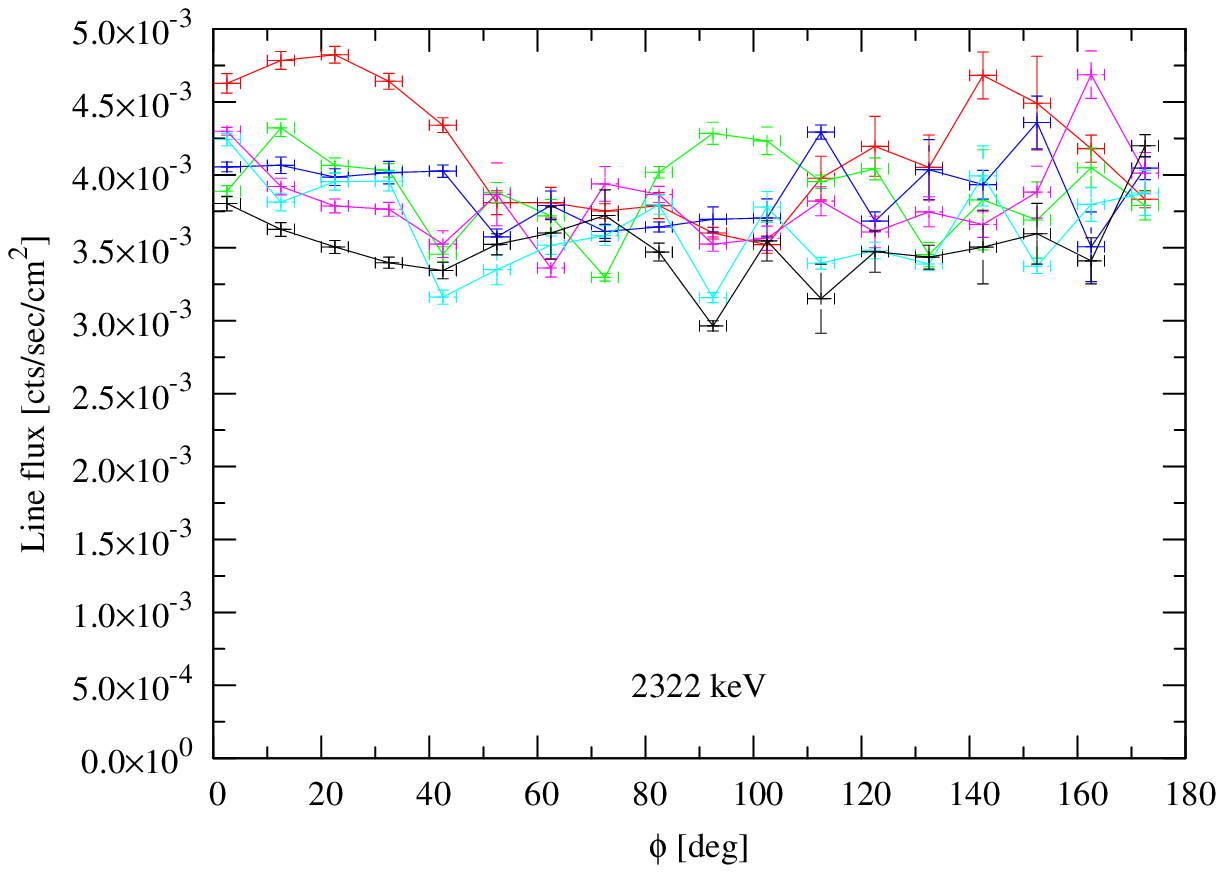}{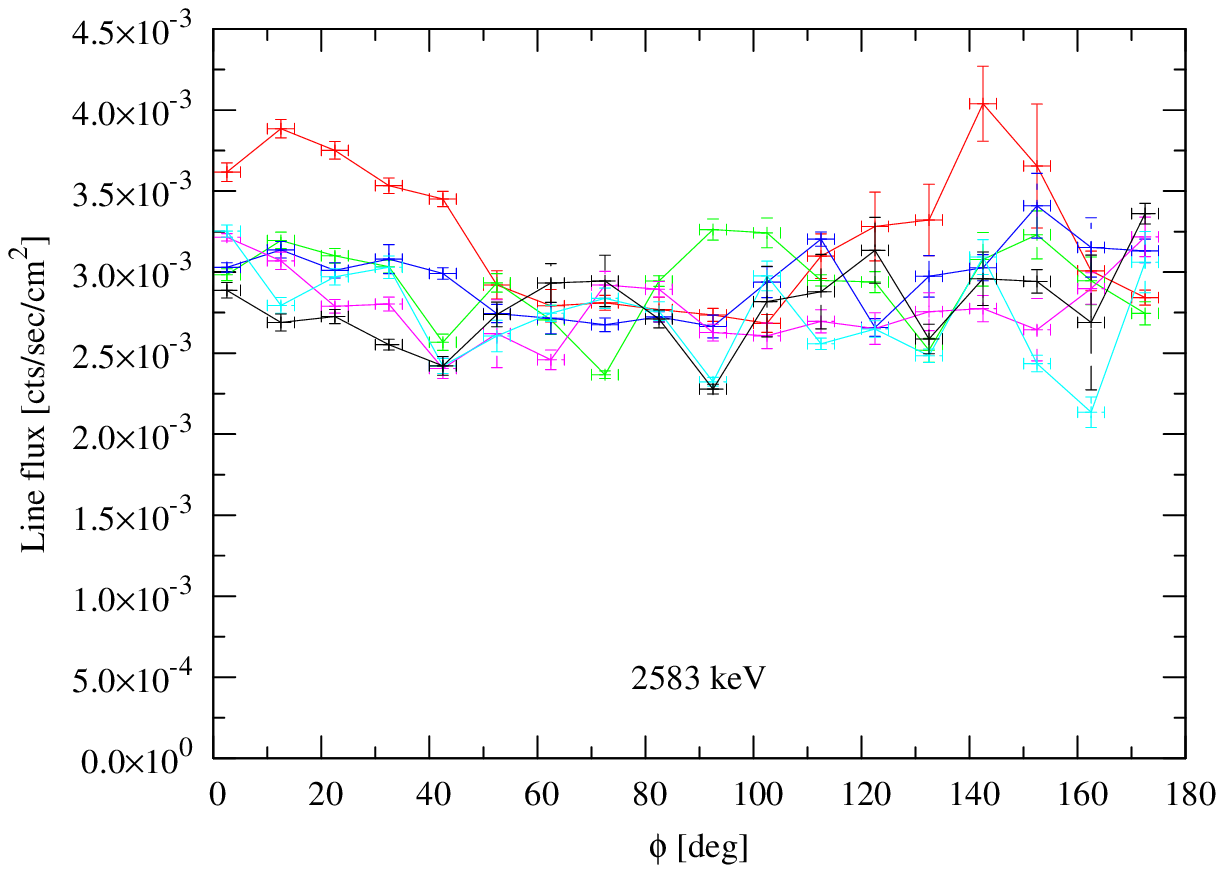}
    \threeLines{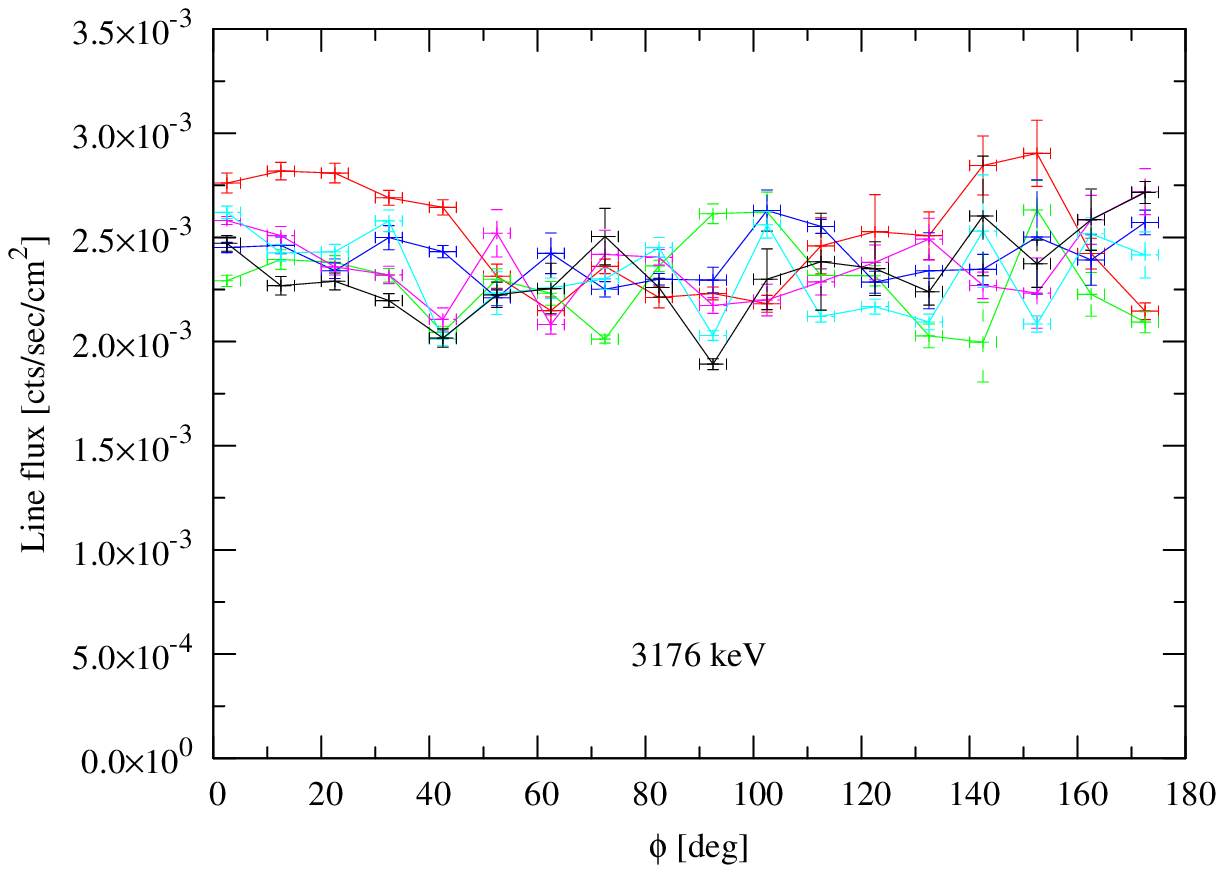}{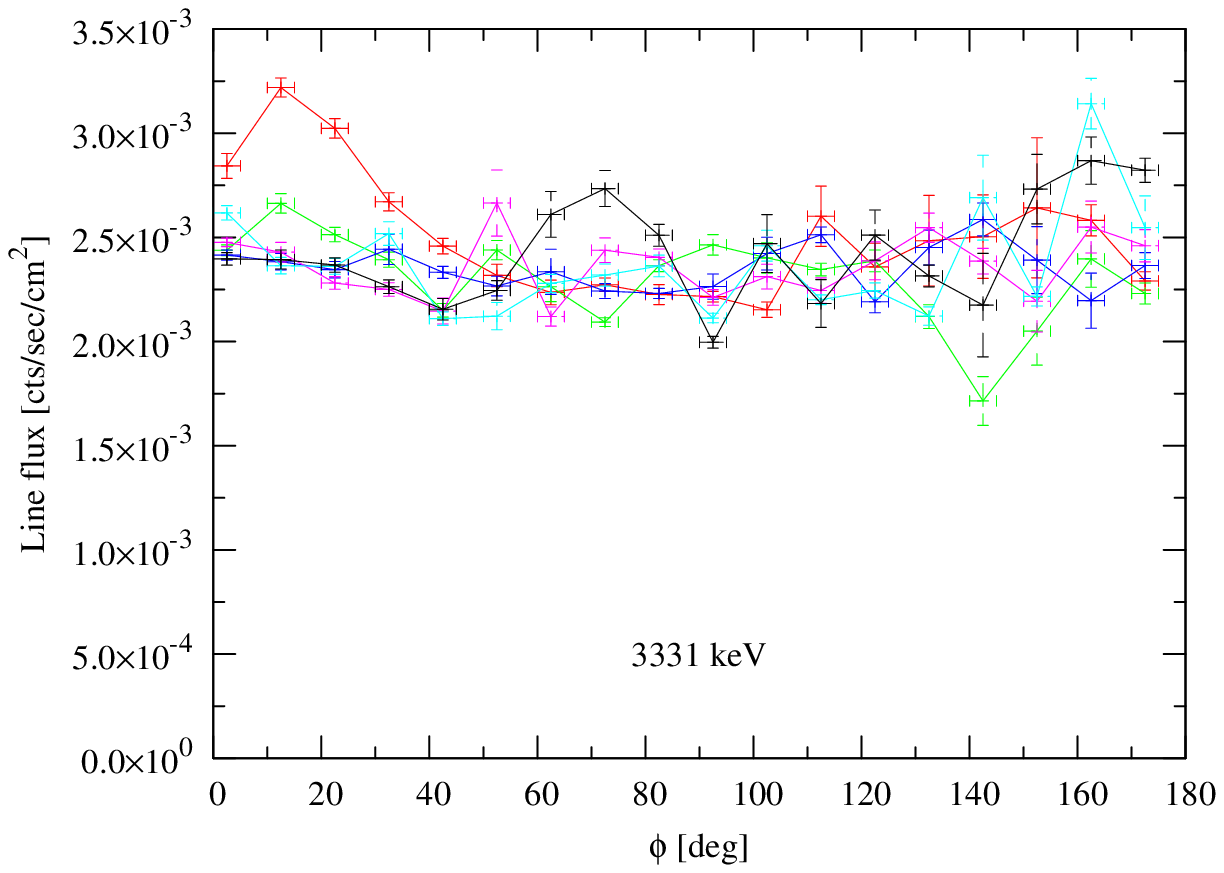}{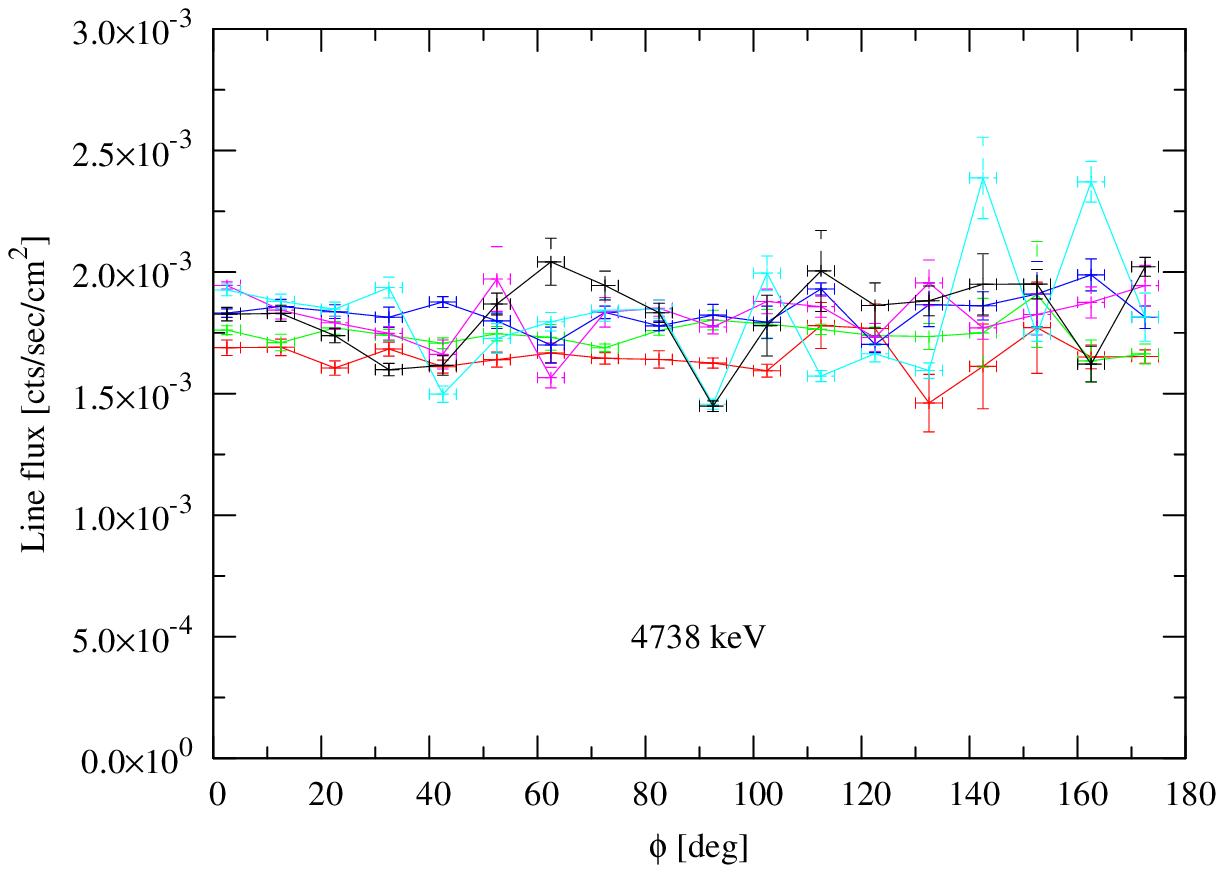}
  \end{tabular}
  \caption{Line flux as a function of the off-GC angle $\phi$ for the
    ``candidate'' lines from Table~\protect\ref{tab:candidates}
    (page~~\protect\pageref{tab:candidates}) . For all lines the flux does not
    depend on the off-GC angle (with $95-100\%$ of all points lying within 3
    standard deviations from the average). Different colors represent
    different phases (see Table~\ref{tab:phases}, p.~~\pageref{tab:phases}).}
  \label{fig:lines}
\end{figure*}

{\footnotesize
\begin{table*}
  \begin{tabular}[c]{rl}
    \begin{minipage}[t]{.5\textwidth}
  \begin{tabular}[t]{|c|c|c|c|c|c|}
    \hline
    E [keV] & $\langle F\rangle $ & $\sigma_F$
    & $F_\text{min}$  & $F_\max$ &
    $\frac{F_\text{max}-F_\text{min}}{F_\min}$\\
    \hline
    330 \input{330.tab}
    385 \input{385.tab}
    431 \input{431.tab}
    511 \input{511.tab}
    937 \input{937.tab}
    947 \input{947.tab}
    1068 \input{1068.tab}
    1098 \input{1098.tab}
    1234 \input{1234.tab}
    1719 \input{1719.tab}
  \end{tabular}
\end{minipage}
&
\begin{minipage}[t]{.5\textwidth}
  \begin{tabular}[t]{|c|c|c|c|c|c|}
    \hline
    E [keV] & $\langle F\rangle $ & $\sigma_F$ & $F_\text{min}$  &
    $F_\text{max}$ &
    $\frac{F_\text{max}-F_\text{min}}{F_\min}$\\
    \hline
    1753 \input{1753.tab}
    1809 \input{1809.tab}
    1904 \input{1904.tab}
    2322 \input{2322.tab}
    2583 \input{2583.tab}
    3176 \input{3176.tab}
    3331 \input{3331.tab}
    4738 \input{4738.tab}
    5186 \input{5186.tab}
    5208 \input{5208.tab}
  \end{tabular}
\end{minipage}\\
\end{tabular}
\caption{Characteristics of the spatial profiles of the candidate
  lines from Table~\protect\ref{tab:candidates}. For each line (and for each of 6 phases) we compute the average
  $\langle F\rangle$, the standard deviation (average scatter of the points
  around its mean value)
  $\sigma_F$, minimal and maximal values and the ratio of
  $(F_\text{max} - F_\text{min})/F_\min$, which gives the \emph{upper} bound on
  the share  of DM, present in the given line. }
  \label{tab:profiles}
\end{table*}}

\label{lastpage}

\end{document}

%% file: line-table-tw120_v3.1.1.tab
68.5 & 11.4 & 0.65 & 66.7 Ge complex\\
76.5 & 58.8 & 1.10 & 75 Bi $K_\alpha$\\
87 & 21.2 & 0.90 & 87 Bi $K_\beta$\\
94 & 5.2 & 0.55 & 91-105 GaZn \\
134.5 & 15.5 & 0.90 & 132-140 Ge complex\\
143 & 12.6 & 1.20 & 140-147 Ge complex\\
177 & 3.9 & 0.95 & 175 AsGe \\
186.5 & 7.3 & 1.10 & 184.6 GaZn \\
193 & 24.8 & 0.75 & 190-198 Ge complex\\
200 & 29.3 & 0.60 & 198-215 Ge complex\\
205.5 & 5.4 & 0.50 & 198-215 Ge complex\\
240 & 3.4 & 1.10 & 238 PbBi \\
302 & 3.9 & 0.70 & 301.5 GaZn \\
311.5 & 7.4 & 1.10 & 309.8 GaZn+K \\
\rowcolor[gray]{.7}
330.5 & 3.2 & 0.55 & 328=? or 331=PbTl\\
\rowcolor[gray]{.7}
385.5 & 3.2 & 1.45 & 383=PbTl or ??\\
404.5 & 5.5 & 0.95 & 403 Ga Zn+K\\
\rowcolor[gray]{.7}
431.5 & 4.4 & 0.55 & ??  \\
440.5 & 12.4 & 0.80 & 438 ZnZn \\
465 & 4.3 & 0.95 & 470-485 NaNa \\
\rowcolor[gray]{.7}
$511^\dagger$ & 52.5 & 1.25 & 511 $e^+e^-$ \\
576 & 5.2 & 0.95 & 574 GeGa \\
585.5 & 7.2 & 1.10 & 584.5 GeGa+K \\
597.5 & 5.0 & 1.45 & 596-610 Ge complex\\
754 & 4.7 & 0.85 & 751 BiBi \\
803.5 & 3.7 & 0.75 & 803 BiPb \\
812 & 10.6 & 0.95 & 810 CoFe \\
819.5 & 11.6 & 0.85 & 817 CoFe+K \\
827.5 & 7.1 & 0.75 & 825 PbPb \\
836 & 12.1 & 0.95 & 834 MnCr \\
845 & 6.5 & 1.30 & 843 MgAl \\
874 & 8.8 & 1.05 & 872 GeGa \\
884 & 10.0 & 1.05 & 882 GeGa+K \\
913 & 5.2 & 1.05 & 911 AcTh \\
\rowcolor[gray]{.7}
$937^\dagger$ & 3.1 & 0.95 & ?? or 935=MnCr\\

%% file: line-table-tw120_v3.1.2.tab
990 & 6.0 & 0.85 & 987 PbPb \\
\rowcolor[gray]{.7}
$947^\dagger$ & 3.6 & 1.50 & ??  \\
1014.5 & 11.1 & 1.40 & 1014 MgAl \\
\rowcolor[gray]{.7}
1068.5 & 3.3 & 1.40 & ?? or 1063=PbPb\\
1079 & 3.3 & 1.00 & 1077=GaZn  \\
\rowcolor[gray]{.7}
1098 & 6.8 & 1.05 & 1095=?  \\
1108.5 & 19.9 & 1.05 & 1106 GeGa \\
1118.5 & 23.9 & 1.05 & 1117 GeGa+K \\
1127 & 23.5 & 0.85 & 1124.5 ZnCu+K \\
\rowcolor[gray]{.7}
$1234^\dagger$ & 4.5 & 1.40 & 1231 TaW \\
1349.5 & 3.5 & 0.80 & 1347 GeGa+K \\
1368.5 & 13.2 & 1.55 & 1368 NaMg \\
\rowcolor[gray]{.7}
$1719.5^\dagger$ & 3.3 & 1.55 & 1719 BiPb \\
\rowcolor[gray]{.7}
1753.5 & 4.0 & 1.45 & ?? or 1758=?\\
1767.5 & 11.1 & 1.40 & 1764 BiPb \\
1781.5 & 12.4 & 1.45 & 1778 AlSi \\
\rowcolor[gray]{.7}
$1809^\dagger$ & 15.2 & 1.85 & 1808 Mg \\
\rowcolor[gray]{.7}
1904 & 3.2 & 1.45 & 1901=GeGa+K or ??\\
2212 & 6.7 & 2.20 & 2195-2223 BiPo, Al\\
2225 & 5.4 & 1.10 & 2223 HD \\
\rowcolor[gray]{.7}
2322 & 3.1 & 1.55 & 2319=?  \\
\rowcolor[gray]{.7}
2583.5 & 3.5 & 1.65 & 2599=? or ??\\
2616 & 5.0 & 1.90 & 2614 PbTl \\
2756 & 9.5 & 1.90 & 2754 NaMg \\
3002.5 & 5.1 & 2.45 & 2993-3013 Al \\
\rowcolor[gray]{.7}
$3176.5^\dagger$ & 3.4 & 1.95 & ??  \\
\rowcolor[gray]{.7}
$3331^\dagger$ & 3.4 & 2.05 & ??  \\
3802 & 3.6 & 2.10 & 3800 GaZn+K \\
4307.5 & 3.6 & 3.40 & 4304 GaZn+K \\
4454 & 4.2 & 8.45 & 4434 C \\
\rowcolor[gray]{.7}
$4738^\dagger$ & 3.5 & 2.30 & ??  \\
\rowcolor[gray]{.7}
$5186.5^\dagger$ & 3.5 & 2.20 & ??  \\
\rowcolor[gray]{.7}
$5208.5\dagger$ & 4.0 & 1.95 & ??  \\
\rowcolor[gray]{.7}
$5757^\dagger$ & 3.4 & 2.95 & ??  \\
6129 & 20.5 & 3.25 & 6128.9 O \\

%% file: 330.tab
 &  9.6e-03 &  5.8e-04 &  8.7e-03 &  1.1e-02 & 0.23  \\
\cline{2-6}
 &  9.6e-03 &  4.0e-04 &  8.9e-03 &  1.0e-02 & 0.15  \\
\cline{2-6}
 &  1.0e-02 &  3.4e-04 &  9.3e-03 &  1.1e-02 & 0.16  \\
\cline{2-6}
 &  1.0e-02 &  5.2e-04 &  9.7e-03 &  1.2e-02 & 0.23  \\
\cline{2-6}
 &  1.0e-02 &  4.7e-04 &  9.6e-03 &  1.2e-02 & 0.24  \\
\cline{2-6}
 &  1.0e-02 &  6.1e-04 &  9.7e-03 &  1.3e-02 & 0.33  \\
\cline{2-6}
\hline

%% file: 385.tab
 &  6.6e-03 &  7.3e-04 &  5.4e-03 &  8.0e-03 & 0.48  \\
\cline{2-6}
 &  6.1e-03 &  4.8e-04 &  5.2e-03 &  6.8e-03 & 0.31  \\
\cline{2-6}
 &  6.3e-03 &  3.9e-04 &  5.3e-03 &  7.1e-03 & 0.32  \\
\cline{2-6}
 &  6.2e-03 &  5.4e-04 &  5.2e-03 &  7.4e-03 & 0.42  \\
\cline{2-6}
 &  6.2e-03 &  5.0e-04 &  5.5e-03 &  7.6e-03 & 0.40  \\
\cline{2-6}
 &  6.7e-03 &  5.5e-04 &  5.8e-03 &  8.4e-03 & 0.44  \\
\cline{2-6}
\hline

%% file: 431.tab
 &  1.1e-02 &  5.3e-04 &  9.2e-03 &  1.2e-02 & 0.26  \\
\cline{2-6}
 &  1.1e-02 &  5.7e-04 &  9.5e-03 &  1.2e-02 & 0.26  \\
\cline{2-6}
 &  1.2e-02 &  1.0e-03 &  1.0e-02 &  1.5e-02 & 0.43  \\
\cline{2-6}
 &  1.3e-02 &  8.1e-04 &  1.1e-02 &  1.4e-02 & 0.32  \\
\cline{2-6}
 &  1.4e-02 &  1.4e-03 &  1.2e-02 &  1.8e-02 & 0.45  \\
\cline{2-6}
 &  1.8e-02 &  1.5e-03 &  1.4e-02 &  2.0e-02 & 0.45  \\
\cline{2-6}
\hline

%% file: 511.tab
 &  4.5e-02 &  7.8e-04 &  4.4e-02 &  4.7e-02 & 0.07  \\
\cline{2-6}
 &  5.9e-02 &  5.8e-04 &  5.8e-02 &  6.0e-02 & 0.04  \\
\cline{2-6}
 &  7.0e-02 &  8.5e-04 &  6.8e-02 &  7.2e-02 & 0.05  \\
\cline{2-6}
 &  7.1e-02 &  6.7e-04 &  6.9e-02 &  7.2e-02 & 0.04  \\
\cline{2-6}
 &  7.3e-02 &  8.7e-04 &  7.1e-02 &  7.4e-02 & 0.05  \\
\cline{2-6}
 &  7.5e-02 &  7.4e-04 &  7.3e-02 &  7.6e-02 & 0.03  \\
\cline{2-6}
\hline

%% file: 937.tab
 &  6.3e-03 &  7.2e-04 &  5.2e-03 &  7.7e-03 & 0.47  \\
\cline{2-6}
 &  5.8e-03 &  5.4e-04 &  4.7e-03 &  6.6e-03 & 0.40  \\
\cline{2-6}
 &  5.7e-03 &  4.3e-04 &  4.9e-03 &  6.5e-03 & 0.33  \\
\cline{2-6}
 &  5.6e-03 &  5.4e-04 &  4.6e-03 &  6.6e-03 & 0.44  \\
\cline{2-6}
 &  5.3e-03 &  5.5e-04 &  4.7e-03 &  7.5e-03 & 0.60  \\
\cline{2-6}
 &  5.5e-03 &  5.0e-04 &  4.6e-03 &  6.6e-03 & 0.44  \\
\cline{2-6}
\hline

%% file: 947.tab
 &  6.0e-03 &  7.2e-04 &  4.9e-03 &  7.4e-03 & 0.51  \\
\cline{2-6}
 &  5.4e-03 &  5.8e-04 &  4.4e-03 &  6.5e-03 & 0.49  \\
\cline{2-6}
 &  5.3e-03 &  4.4e-04 &  4.4e-03 &  6.1e-03 & 0.40  \\
\cline{2-6}
 &  5.2e-03 &  5.0e-04 &  4.3e-03 &  6.1e-03 & 0.43  \\
\cline{2-6}
 &  5.0e-03 &  6.0e-04 &  4.3e-03 &  7.0e-03 & 0.61  \\
\cline{2-6}
 &  5.3e-03 &  5.2e-04 &  4.2e-03 &  6.5e-03 & 0.55  \\
\cline{2-6}
\hline

%% file: 1068.tab
 &  6.9e-03 &  7.2e-04 &  5.8e-03 &  8.2e-03 & 0.40  \\
\cline{2-6}
 &  6.3e-03 &  5.5e-04 &  5.2e-03 &  7.3e-03 & 0.41  \\
\cline{2-6}
 &  6.1e-03 &  4.2e-04 &  5.3e-03 &  7.0e-03 & 0.33  \\
\cline{2-6}
 &  6.0e-03 &  5.3e-04 &  5.1e-03 &  7.2e-03 & 0.41  \\
\cline{2-6}
 &  5.6e-03 &  4.5e-04 &  4.9e-03 &  6.8e-03 & 0.38  \\
\cline{2-6}
 &  5.5e-03 &  4.1e-04 &  4.6e-03 &  6.6e-03 & 0.43  \\
\cline{2-6}
\hline

%% file: 1098.tab
 &  3.7e-03 &  3.9e-04 &  3.1e-03 &  4.6e-03 & 0.49  \\
\cline{2-6}
 &  3.7e-03 &  2.5e-04 &  3.1e-03 &  4.0e-03 & 0.30  \\
\cline{2-6}
 &  4.5e-03 &  6.6e-04 &  3.1e-03 &  6.1e-03 & 0.95  \\
\cline{2-6}
 &  4.5e-03 &  4.8e-04 &  3.6e-03 &  5.5e-03 & 0.56  \\
\cline{2-6}
 &  5.4e-03 &  1.0e-03 &  3.8e-03 &  7.6e-03 & 0.99  \\
\cline{2-6}
 &  7.6e-03 &  1.3e-03 &  5.1e-03 &  9.9e-03 & 0.94  \\
\cline{2-6}
\hline

%% file: 1234.tab
 &  6.4e-03 &  5.9e-04 &  5.4e-03 &  7.4e-03 & 0.36  \\
\cline{2-6}
 &  6.2e-03 &  5.4e-04 &  5.2e-03 &  7.2e-03 & 0.38  \\
\cline{2-6}
 &  6.0e-03 &  3.7e-04 &  5.2e-03 &  6.6e-03 & 0.27  \\
\cline{2-6}
 &  6.1e-03 &  4.6e-04 &  5.4e-03 &  6.9e-03 & 0.27  \\
\cline{2-6}
 &  5.8e-03 &  5.5e-04 &  5.1e-03 &  7.6e-03 & 0.50  \\
\cline{2-6}
 &  5.8e-03 &  4.0e-04 &  5.0e-03 &  6.8e-03 & 0.35  \\
\cline{2-6}
\hline

%% file: 1719.tab
 &  4.9e-03 &  5.0e-04 &  4.0e-03 &  5.7e-03 & 0.42  \\
\cline{2-6}
 &  4.6e-03 &  3.8e-04 &  3.8e-03 &  5.1e-03 & 0.35  \\
\cline{2-6}
 &  4.7e-03 &  3.1e-04 &  3.9e-03 &  5.2e-03 & 0.34  \\
\cline{2-6}
 &  4.6e-03 &  3.8e-04 &  4.1e-03 &  5.4e-03 & 0.32  \\
\cline{2-6}
 &  4.5e-03 &  4.0e-04 &  3.9e-03 &  5.5e-03 & 0.40  \\
\cline{2-6}
 &  4.7e-03 &  4.2e-04 &  3.9e-03 &  5.7e-03 & 0.46  \\
\cline{2-6}
\hline

%% file: 1753.tab
 &  3.2e-03 &  4.1e-04 &  2.6e-03 &  4.0e-03 & 0.53  \\
\cline{2-6}
 &  3.0e-03 &  2.7e-04 &  2.4e-03 &  3.4e-03 & 0.41  \\
\cline{2-6}
 &  3.1e-03 &  2.7e-04 &  2.4e-03 &  3.7e-03 & 0.59  \\
\cline{2-6}
 &  3.1e-03 &  3.0e-04 &  2.6e-03 &  3.7e-03 & 0.43  \\
\cline{2-6}
 &  3.1e-03 &  3.5e-04 &  2.6e-03 &  3.9e-03 & 0.51  \\
\cline{2-6}
 &  3.6e-03 &  4.5e-04 &  2.7e-03 &  4.3e-03 & 0.61  \\
\cline{2-6}
\hline

%% file: 1809.tab
 &  5.5e-03 &  5.1e-04 &  4.6e-03 &  6.6e-03 & 0.43  \\
\cline{2-6}
 &  5.2e-03 &  3.3e-04 &  4.6e-03 &  5.8e-03 & 0.28  \\
\cline{2-6}
 &  5.7e-03 &  3.3e-04 &  5.0e-03 &  6.3e-03 & 0.27  \\
\cline{2-6}
 &  5.7e-03 &  3.5e-04 &  5.2e-03 &  6.5e-03 & 0.24  \\
\cline{2-6}
 &  5.9e-03 &  4.7e-04 &  5.0e-03 &  7.3e-03 & 0.46  \\
\cline{2-6}
 &  6.1e-03 &  3.9e-04 &  5.4e-03 &  7.0e-03 & 0.30  \\
\cline{2-6}
\hline

%% file: 1904.tab
 &  4.8e-03 &  5.1e-04 &  4.1e-03 &  5.6e-03 & 0.39  \\
\cline{2-6}
 &  4.5e-03 &  3.9e-04 &  3.8e-03 &  5.3e-03 & 0.40  \\
\cline{2-6}
 &  4.4e-03 &  3.2e-04 &  3.7e-03 &  5.2e-03 & 0.39  \\
\cline{2-6}
 &  4.3e-03 &  4.0e-04 &  3.6e-03 &  5.2e-03 & 0.44  \\
\cline{2-6}
 &  4.1e-03 &  3.7e-04 &  3.6e-03 &  5.2e-03 & 0.42  \\
\cline{2-6}
 &  4.1e-03 &  3.3e-04 &  3.4e-03 &  4.9e-03 & 0.43  \\
\cline{2-6}
\hline

%% file: 2322.tab
 &  4.2e-03 &  4.0e-04 &  3.5e-03 &  4.9e-03 & 0.39  \\
\cline{2-6}
 &  3.9e-03 &  3.0e-04 &  3.2e-03 &  4.3e-03 & 0.34  \\
\cline{2-6}
 &  3.9e-03 &  2.9e-04 &  3.3e-03 &  4.7e-03 & 0.42  \\
\cline{2-6}
 &  3.8e-03 &  3.2e-04 &  3.3e-03 &  4.7e-03 & 0.43  \\
\cline{2-6}
 &  3.6e-03 &  2.9e-04 &  3.1e-03 &  4.2e-03 & 0.35  \\
\cline{2-6}
 &  3.5e-03 &  2.8e-04 &  3.0e-03 &  4.2e-03 & 0.42  \\
\cline{2-6}
\hline

%% file: 2583.tab
 &  3.2e-03 &  4.1e-04 &  2.6e-03 &  4.0e-03 & 0.53  \\
\cline{2-6}
 &  2.9e-03 &  2.6e-04 &  2.4e-03 &  3.3e-03 & 0.41  \\
\cline{2-6}
 &  2.9e-03 &  2.2e-04 &  2.4e-03 &  3.4e-03 & 0.42  \\
\cline{2-6}
 &  2.8e-03 &  2.6e-04 &  2.4e-03 &  3.4e-03 & 0.44  \\
\cline{2-6}
 &  2.7e-03 &  2.6e-04 &  2.1e-03 &  3.3e-03 & 0.52  \\
\cline{2-6}
 &  2.8e-03 &  2.6e-04 &  2.3e-03 &  3.4e-03 & 0.48  \\
\cline{2-6}
\hline

%% file: 3176.tab
 &  2.5e-03 &  2.5e-04 &  2.1e-03 &  3.0e-03 & 0.41  \\
\cline{2-6}
 &  2.3e-03 &  1.9e-04 &  1.9e-03 &  2.7e-03 & 0.40  \\
\cline{2-6}
 &  2.4e-03 &  1.6e-04 &  1.9e-03 &  2.7e-03 & 0.41  \\
\cline{2-6}
 &  2.4e-03 &  1.9e-04 &  2.0e-03 &  2.7e-03 & 0.36  \\
\cline{2-6}
 &  2.3e-03 &  2.1e-04 &  1.9e-03 &  2.7e-03 & 0.42  \\
\cline{2-6}
 &  2.3e-03 &  2.2e-04 &  1.9e-03 &  2.9e-03 & 0.54  \\
\cline{2-6}
\hline

%% file: 3331.tab
 &  2.5e-03 &  2.9e-04 &  2.2e-03 &  3.2e-03 & 0.50  \\
\cline{2-6}
 &  2.3e-03 &  2.0e-04 &  1.7e-03 &  2.7e-03 & 0.55  \\
\cline{2-6}
 &  2.4e-03 &  1.5e-04 &  2.0e-03 &  2.8e-03 & 0.35  \\
\cline{2-6}
 &  2.3e-03 &  1.7e-04 &  2.0e-03 &  2.7e-03 & 0.33  \\
\cline{2-6}
 &  2.3e-03 &  2.1e-04 &  2.1e-03 &  3.1e-03 & 0.51  \\
\cline{2-6}
 &  2.4e-03 &  2.3e-04 &  2.0e-03 &  2.9e-03 & 0.44  \\
\cline{2-6}
\hline

%% file: 4738.tab
 &  1.7e-03 &  6.2e-05 &  1.5e-03 &  1.8e-03 & 0.22  \\
\cline{2-6}
 &  1.7e-03 &  6.5e-05 &  1.5e-03 &  1.9e-03 & 0.24  \\
\cline{2-6}
 &  1.8e-03 &  9.9e-05 &  1.5e-03 &  2.0e-03 & 0.31  \\
\cline{2-6}
 &  1.8e-03 &  1.3e-04 &  1.5e-03 &  2.0e-03 & 0.38  \\
\cline{2-6}
 &  1.8e-03 &  2.1e-04 &  1.5e-03 &  2.4e-03 & 0.64  \\
\cline{2-6}
 &  1.8e-03 &  1.7e-04 &  1.4e-03 &  2.1e-03 & 0.46  \\
\cline{2-6}
\hline

%% file: 5186.tab
 &  1.5e-03 &  6.7e-05 &  1.4e-03 &  1.7e-03 & 0.26  \\
\cline{2-6}
 &  1.6e-03 &  5.5e-05 &  1.5e-03 &  1.8e-03 & 0.16  \\
\cline{2-6}
 &  1.8e-03 &  9.4e-05 &  1.6e-03 &  2.1e-03 & 0.35  \\
\cline{2-6}
 &  1.8e-03 &  6.4e-05 &  1.6e-03 &  2.0e-03 & 0.24  \\
\cline{2-6}
 &  1.8e-03 &  6.9e-05 &  1.6e-03 &  1.9e-03 & 0.18  \\
\cline{2-6}
 &  1.8e-03 &  1.2e-04 &  1.5e-03 &  2.0e-03 & 0.32  \\
\cline{2-6}
\hline

%% file: 5208.tab
 &  1.5e-03 &  8.8e-05 &  1.2e-03 &  1.8e-03 & 0.51  \\
\cline{2-6}
 &  1.6e-03 &  5.6e-05 &  1.5e-03 &  1.8e-03 & 0.21  \\
\cline{2-6}
 &  1.7e-03 &  5.8e-05 &  1.6e-03 &  1.9e-03 & 0.16  \\
\cline{2-6}
 &  1.7e-03 &  5.3e-05 &  1.6e-03 &  1.9e-03 & 0.15  \\
\cline{2-6}
 &  1.8e-03 &  8.0e-05 &  1.6e-03 &  2.1e-03 & 0.28  \\
\cline{2-6}
 &  1.8e-03 &  1.3e-04 &  1.5e-03 &  2.1e-03 & 0.39  \\
\cline{2-6}
\hline